\documentclass[journal=jctcce,manuscript=article,layout=twocolumn]{achemso}
\usepackage[version=3]{mhchem}
\usepackage{amssymb}
\usepackage{color}
\usepackage{braket}
\usepackage{xspace}
\usepackage{cleveref}
\usepackage{graphicx}
\usepackage{subfig}
\usepackage{threeparttable}
\usepackage{textcomp}

\newcommand*{\eh}{\ensuremath{E_\mathrm{h}}\xspace}
\newcommand*{\mae}{$\Delta_{\mathrm{MAE}}$\xspace}
\newcommand*{\std}{$\Delta_{\mathrm{STD}}$\xspace}
\newcommand*{\maxe}{$\Delta_{\mathrm{MAX}}$\xspace}
\newcommand*{\kcal}{kcal mol$^{-1}$\xspace} 
\newcommand*{\au}{\ensuremath{\mathrm{a.u.}}\xspace}

\crefname{figure}{Figure}{Figures}      
\crefname{table}{Table}{Tables}         
\crefname{equation}{Eq.}{Eqs.}          
\crefname{section}{Section}{Sections}   
\crefname{subsection}{Section}{Sections}

\author{Andreas V.\ Copan}
\email{avcopan@gmail.com}
\affiliation{%
     Department of Chemistry and Biochemistry,
     The Ohio State University,
     Columbus, Ohio 43210, United States
}
 \author{Alexander Yu.\ Sokolov}
 \email{sokolov.8@osu.edu}
 \affiliation{%
     Department of Chemistry and Biochemistry,
     The Ohio State University,
     Columbus, Ohio 43210, United States
 }

\begin{tocentry}
\includegraphics{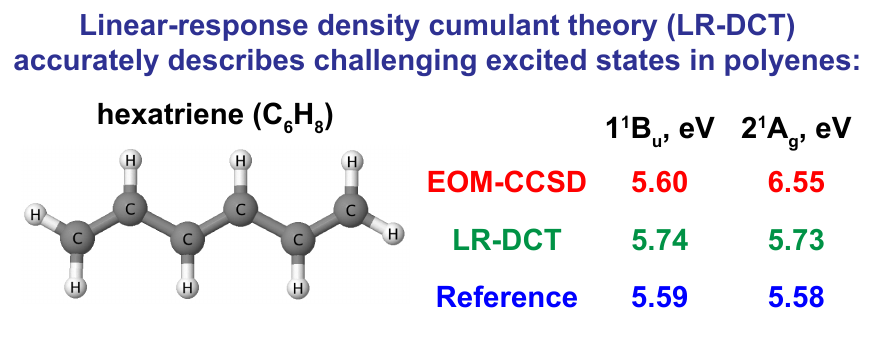}
\end{tocentry}

\title{Linear-response density cumulant theory for excited electronic states}

\begin{document}

\abstract{%
We present a linear-response formulation of density cumulant theory (DCT) that
provides a balanced and accurate description of many electronic states
simultaneously.
In the original DCT formulation, only information about a single electronic
state (usually, the ground state) is obtained.
We discuss the derivation of linear-response DCT, present its implementation for
the ODC-12 method (LR-ODC-12), and benchmark its performance for excitation
energies in small molecules (\ce{N2}, \ce{CO}, \ce{HCN}, \ce{HNC}, \ce{C2H2},
and \ce{H2CO}), as well as challenging excited states in ethylene, butadiene,
and hexatriene. 
For small molecules, LR-ODC-12 shows smaller mean absolute errors in excitation energies than equation-of-motion coupled cluster
theory with single and double excitations (EOM-CCSD), relative to the reference
data from EOM-CCSDT\@.
In a study of butadiene and hexatriene, LR-ODC-12 correctly describes the
relative energies of the singly-excited $1{}^1\mathrm{B_{u}}$ and the
doubly-excited $2{}^1\mathrm{A_{g}}$ states, in excellent agreement with
highly accurate semistochastic heat-bath configuration interaction results,
while EOM-CCSD overestimates the energy of the $2{}^1\mathrm{A_{g}}$ state by
almost 1 eV.
Our results demonstrate that linear-response DCT is a promising theoretical approach for excited states of molecules. 
}

\section{Introduction}
Accurate simulation of excited electronic states remains one of the major
challenges in modern electronic structure theory. 
{\it Ab initio}\/ methods for excited states can be divided into
single-reference and multi-reference categories, based on their ability to
treat static electron correlation.
Multi-reference methods 
\cite{Knowles:1985p259,Wolinski:1987p225,Hirao:1992p374,Finley:1998p299,Andersson:1990p5483,Andersson:1992p1218,Angeli:2001p10252,Angeli:2001p297,Mukherjee:1977p955,Jeziorski:1981p1668,Werner:1988p5803,Mahapatra:1998p157,Pittner:2003p10876,Evangelista:2007p024102,Datta:2011p214116,Evangelista:2011p114102,Kohn:2012p176,Nooijen:2014p081102}
can correctly describe static correlation in
near-degenerate valence orbitals and electronic states with multiple-excitation
character, but often lack accurate treatment of important dynamic correlation
effects
or become computationally very costly when the number of strongly correlated orbitals is large.
Meanwhile,
single-reference methods
\cite{Foresman:1992p135,Sherrill:1999p143,Geertsen:1989p57,Comeau:1993p414,Stanton:1993p7029,Krylov:2008p433,Crawford:2000p33,Shavitt:2009,Sekino:1984p255,Koch:1990p3345,Koch:1990p3333,Nooijen:1997p6441,Nooijen:1997p6812,Nakatsuji:1978p2053,Nakatsuji:1979p329}
often provide a compromise between the computational
cost and accuracy, and can be used to reliably compute properties of molecules
in low-lying electronic states near the equilibrium geometries. In these
situations, single-reference equation-of-motion coupled cluster theory
(EOM-CC)
\cite{Geertsen:1989p57,Comeau:1993p414,Stanton:1993p7029,Krylov:2008p433,Crawford:2000p33,Shavitt:2009}
is usually the method of choice, especially when high accuracy
is desired. 

The EOM-CC methods yield size-intensive excitation energies
\cite{Koch:1990p3345,Koch:1990p3333}
and can be
systematically improved by increasing the excitation rank of the cluster
operator in the exponential parametrization of the wavefunction. Although EOM-CC
is usually formulated in the context of a similarity-transformed Hamiltonian,
its excitation energies are equivalent to those obtained from linear-response
coupled cluster theory (LR-CC).
\cite{Sekino:1984p255,Koch:1990p3345,Koch:1990p3333}
Both EOM-CC and LR-CC are based on 
non-Hermitian eigenvalue problems,
which complicates the computation of molecular properties (e.g., transition dipoles)
by requiring evaluation of left and right eigenvectors, 
\cite{Stanton:1993p8840,Stanton:1994p4695,Stanton:1994p8938,Levchenko:2005p224106}
and may result in an incorrect description of potential energy surfaces in the
vicinity of conical intersections where complex excitation energies may be
obtained.\cite{Hattig:2005p37,Kohn:2007p044105,Kjonstad:2017p164105}
Several Hermitian alternatives to EOM-CC and LR-CC have been
proposed to avoid these problems, such as  
algebraic diagrammatic construction
\cite{Schirmer:1982p2395,Schirmer:1991p4647,Dreuw:2014p82}, 
unitary and variational LR-CC,
\cite{Taube:2006p3393,Kats:2011p062503,Walz:2012p052519}
similarity-constrained CC,
\cite{Kjonstad:2017p4801}
and propagator-based LR-CC.
\cite{Moszynski:2005p1109,Korona:2010p14977}

In this work, we present a linear-response formulation of density cumulant
theory for excited electronic states.
In density cumulant theory (DCT),
\cite{Kutzelnigg:2006p171101,Simmonett:2010p174122,Sokolov:2012p054105,Sokolov:2013p024107,Sokolov:2013p204110,Sokolov:2014p074111,Wang:2016p4833,DCFTfootnote}
the electronic energy is determined directly in terms of the one-particle
reduced density matrix and the density
cumulant, i.e.~the fully connected part of the two-body reduced density matrix
(2-RDM).
\cite{Fulde:1991,Ziesche:1992p597,Kutzelnigg:1997p432,Mazziotti:1998p419,Mazziotti:1998p4219,Kutzelnigg:1999p2800,Ziesche:2000p33,Herbert:2007p261,Kong:2011p214109,Hanauer:2012p50}
In this regard, DCT is related to approaches based on the variational
optimization
\cite{Colmenero:1993p979,Nakatsuji:1996p1039,Mazziotti:1998p4219,Nakata:2001p8282,Nakata:2002p5432,Mazziotti:2006p143002,Kollmar:2006p084108,DePrince:2007p042501,DePrince:2016p164109}
or parametrization
\cite{Mazziotti:2008p253002,Mazziotti:2010p062515,DePrince:2012p1917} of the
2-RDM\@.
On the other hand, DCT has a close relationship with wavefunction-based
electronic structure theories, 
\cite{Sokolov:2013p024107,Sokolov:2013p204110}
such as linearized, unitary, and variational coupled
cluster theory.
\cite{Kutzelnigg:1991p349,Kutzelnigg:1998p65,VanVoorhis:2000p8873,Kutzelnigg:1982p3081,Bartlett:1989p133,Watts:1989p359,Szalay:1995p281,Cooper:2010p234102,Evangelista:2011p224102}
In contrast to variational 2-RDM theory
\cite{Nakata:2009p042109,vanAggelen:2010p114112,Verstichel:2010p114113}
and traditional
coupled cluster methods,\cite{Crawford:2000p33,Shavitt:2009}
DCT naturally combines
size-extensivity and a Hermitian energy functional. 
In addition, the DCT
electronic energy is fully optimized with respect to all of its parameters, which
greatly simplifies computation of the first-order molecular properties.
\cite{Scheiner:1987p5361,Salter:1989p1752,Gauss:1991p2623,Gauss:1991p207}
We have successfully applied DCT to a variety of chemical systems with
different electronic structure effects (e.g., open-shell, symmetry-breaking,
and multi-reference).
\cite{Sokolov:2013p204110,Sokolov:2014p074111,Wang:2016p4833,Copan:2014p2389,Mullinax:2015p2487}
One limitation of the original DCT formulation is
the ability to describe only the lowest-energy state of a
particular symmetry (usually, the ground state). By combining DCT with linear
response theory, we remove this limitation, providing access to many electronic
states simultaneously.

We begin with a brief overview of DCT (\cref{sec:dct}) and linear response
theory (\cref{sec:lr}).
In \cref{sec:lr_odc12}, we describe the derivation of the linear-response
equations for the ODC-12 model (LR-ODC-12).
In \cref{sec:olccd}, we compare the LR-ODC-12 method with linear-response
orbital-optimized linearized coupled cluster theory with double excitations
(LR-OLCCD), which we derive by linearizing the LR-ODC-12 equations.
We outline the
computational details in \cref{sec:comp_details}.
In \cref{sec:results}, we demonstrate that the LR-ODC-12 excitation
energies are size-intensive (\cref{sec:size_intensivity}), test the performance
of LR-ODC-12 for the dissociation of \ce{H2} (\cref{sec:two_electron}),
benchmark its accuracy for vertical
excitation energies of small molecules (\cref{sec:vert_excit}), and apply LR-ODC-12 to challenging excited states in ethylene, butadiene, and hexatriene (\cref{sec:alkenes}).
We present our conclusions in \cref{sec:conclusions}. 

\section{Theory}

\subsection{Overview of Density Cumulant Functional Theory}
\label{sec:dct}

We begin with a brief overview of density cumulant theory (DCT) for a single
electronic state.
Our starting point is to express the electronic energy as a trace of the one-
and antisymmetrized two-electron integrals (\( h_p^q \) and
\(\overline{g}_{pq}^{rs}\)) with the reduced one- and two-body density matrices
(\(\gamma^p_q\) and \(\gamma^{pq}_{rs}\)):
\begin{equation}
    \label{eq:energy-expression}
    E
    =
    h_p^q
    \gamma^p_q
    +
    \tfrac{1}{4}
    \overline{g}_{pq}^{rs}
    \gamma^{pq}_{rs}
\end{equation}
where summation over the repeated indices is implied.
In DCT, the two-body density matrix \(\gamma^{pq}_{rs}\) is expanded in terms of
its connected part, the two-body density cumulant ($\lambda^{pq}_{rs}$), and its
disconnected part, which is given by an antisymmetrized product of one-body
density matrices:\cite{Kutzelnigg:2006p171101}
\begin{equation}
    \label{eq:two-body-n-rep}
    \gamma^{pq}_{rs}
    =
    \langle\Psi|
    a^{pq}_{rs}
    |\Psi\rangle
    =
    \lambda^{pq}_{rs}
    +
    P_{(r/s)}
    \gamma^p_r
    \gamma^q_s
\end{equation}
where \(P_{(r/s)}v_{rs} = v_{rs} - v_{sr}\) denotes antisymmetrization and
\mbox{$a^{pq}_{rs}=a^{\dag}_{p}a^{\dag}_{q}a^{}_{s}a^{}_{r}$} is the two-body operator in second quantization.
The one-body density matrix \(\gamma^p_q\) is determined from its non-linear
relationship to the cumulant's partial trace:\cite{Sokolov:2013p024107}
\begin{equation}
    \label{eq:one-body-n-rep}
    \gamma^p_q
    =
    \gamma^p_r
    \gamma^r_q
    -
    \lambda^{pr}_{qr}
\end{equation}
This allows us to determine the energy \eqref{eq:energy-expression} from the
two-body density cumulant and the spin-orbitals, thereby defining the DCT energy
functional.
The density cumulant is parametrized by choosing a specific Ansatz for the
wavefunction \(|\Psi\rangle\) such that\cite{Sokolov:2014p074111}
\begin{equation}
    \label{eq:cumulant-parametrization}
    \lambda^{pq}_{rs}
    =
    \langle\Psi|
    a^{pq}_{rs}
    |\Psi\rangle_c
\end{equation}
where $c$ indicates that only fully connected terms are included in the
parametrization.
Importantly, due to the connected nature of \cref{eq:cumulant-parametrization},
DCT is both size-consistent and size-extensive for any parametrization of
\(|\Psi\rangle\), and is exact in the limit of a complete
parametrization (i.e., when $|\Psi\rangle$ is expanded in the full Hilbert space).\cite{Sokolov:2014p074111}
\cref{eq:cumulant-parametrization} can be considered as a set of
\(n\)-representability conditions that constrain the resulting one- and two-body density matrices to (at least approximately) represent a physical \(n\)-electron wavefunction. 
To compute the DCT energy, the functional \eqref{eq:energy-expression} is made
stationary with respect to all of its parameters.

In this work, we consider the ODC-12
method,\cite{Sokolov:2013p024107,Sokolov:2013p204110} which uses an approximate parametrization for the 
cumulant in \cref{eq:cumulant-parametrization} where the wavefunction $\ket{\Psi}$ is expressed using a unitary transformation truncated at the second order:\cite{Sokolov:2014p074111}
\begin{align}
    \label{eq:cumulant-parametrization_approx}
    \lambda^{pq}_{rs}
    &\approx
    \langle\Phi| e^{-(\hat{T}^{}_2-\hat{T}^\dag_2)} a^{pq}_{rs} e^{\hat{T}^{}_2-\hat{T}^\dag_2} |\Phi\rangle_c \notag \\
    &\approx
    \langle\Phi| a^{pq}_{rs}|\Phi\rangle_c
    + \langle\Phi| [a^{pq}_{rs},\hat{T}^{}_2-\hat{T}^\dag_2]|\Phi\rangle_c \notag \\
    &+ \frac{1}{2}\langle\Phi| [[a^{pq}_{rs},\hat{T}^{}_2-\hat{T}^\dag_2], \hat{T}^{}_2-\hat{T}^\dag_2]|\Phi\rangle_c
\end{align}
\begin{equation}
    \hat{T}_2
    =
    \mathbf{t}_2\cdot\mathbf{a}_2
    =
    \tfrac{1}{4}
    t_{ab}^{ij}
    a^{ab}_{ij}
\end{equation}
Approximation in \cref{eq:cumulant-parametrization_approx} is equivalent to choosing an approximate form for the wavefunction,
\begin{equation}
    \label{eq:odc12-wavefunction}
    |\Psi\rangle
    \approx
    e^{\hat{T}_1-\hat{T}_1^\dagger}
    (1 + \hat{T}_2)
    |\Phi\rangle
\end{equation}
\begin{equation}
    \hat{T}_1
    =
    \mathbf{t}_1\cdot\mathbf{a}_1
    =
    t_a^i
    a^a_i
\end{equation}
inserting it in \cref{eq:cumulant-parametrization}, and keeping only the connected contributions. In \cref{eq:odc12-wavefunction}, we have additionally introduced the unitary singles operator \(e^{\hat{T}_1-\hat{T}_1^\dagger}\) that incorporates orbital relaxation. In our ODC-12 implementation, the effect of the \(e^{\hat{T}_1-\hat{T}_1^\dagger}\) operator is included by optimizing the
orbitals.\cite{Sokolov:2013p204110} We note that, since the unitary transformation in \cref{eq:cumulant-parametrization_approx} is truncated at the second order, the ODC-12 method is not exact for two-electron systems, although the wavefunction in \cref{eq:odc12-wavefunction} is exact for two-electron systems.
The \(\mathbf{t}_1\) and \(\mathbf{t}_2\) parameters are obtained from the
stationarity conditions
\begin{equation}
    \label{eq:stationarity_conditions}
    \dfrac{\partial E}{\partial \mathbf{t}_1^\dagger}
    \overset{!}{=}
    0 \ ,
    \qquad
    \dfrac{\partial E}{\partial \mathbf{t}_2^\dagger}
    \overset{!}{=}
    0
\end{equation}
and are used to compute the ODC-12 energy.
Explicit equations for the stationarity conditions are given in Refs.\@
\citenum{Sokolov:2013p024107} and \citenum{Sokolov:2013p204110}.
Although in ODC-12 the wavefunction parametrization is linear with respect
to double excitations (\cref{eq:odc12-wavefunction}), the ODC-12 energy
stationarity conditions are non-linear in $\mathbf{t}_2$ due to the non-linear
relationship between the one-particle density matrix and the density cumulant
(\cref{eq:one-body-n-rep}).\cite{Sokolov:2013p024107} Neglecting the non-linear
$\mathbf{t}_2$ terms in \cref{eq:stationarity_conditions} results in the
equations that define the linearized orbital-optimized  coupled cluster doubles
method (OLCCD).
This method is equivalent to the orbital-optimized coupled electron pair
approximation zero (OCEPA$_0$).\cite{Bozkaya:2013p054104}

\subsection{Linear Response Theory}
\label{sec:lr}

We now briefly review linear response theory in the quasi-energy
formulation.\cite{Norman:2011p20519}
For a more detailed presentation, we refer the readers to
Ref.~\citenum{Helgaker:2012p543}.
The quasi-energy of a system perturbed by a time-dependent interaction
\(\hat{V}f(t)\) is defined as
\begin{equation}
    Q(t)
    =
    \langle\Psi(t)|
    \hat{H} + \hat{V} f(t) - i\tfrac{\partial}{\partial t}
    |\Psi(t)\rangle
\end{equation}
where \(\Psi(t)\) is the phase-isolated wavefunction,
from which the usual Schr\"odinger wavefunction can be recovered as follows:
\begin{equation}
    \Psi_\mathrm{S}(t)
    =
    e^{-i\int_0^t dt' Q(t')}
    \Psi(t)
\end{equation}
Assuming that the perturbation is Hermitian and periodic, the time average of
the quasi-energy over a period of oscillation, denoted as \( \{Q(t)\} \), is
variational with respect to the exact dynamic state.\cite{Helgaker:2012p543}
The time-dependence of the perturbation can be expressed as a Fourier
expansion
\begin{equation}
    f(t)
    =
    \sum_\omega f(\omega) e^{-i\omega t}
\end{equation}
where the sum runs over frequencies of a common period, and Hermiticity demands
that the negative frequencies are included as well to satisfy the condition
\(f(-\omega)=f^*(\omega)\).
The independent parameters \(\mathbf{u}(t)\) defining the time-dependent
wavefunction can be expressed in polynomial orders of \(f(t)\) as
\begin{equation}
    \label{eq:parameter-fourier-expansion}
    \mathbf{u}(t)
    =
    \mathbf{u}
    +
    \sum_{\omega}
    \mathbf{u}(\omega)
    e^{-i\omega t}
    +
    \cdots
\end{equation}
where only the linear (first-order) contribution is relevant in the present work.
The stationarity of the time-averaged quasi-energy implies the following
relationship\cite{Kristensen:2009p044112}
\begin{equation}
    \label{eq:linear-response-equation}
    \begin{array}{l}
        0
        =
        \left.
            \dfrac{d}{df(\omega)}
            \dfrac{%
                \partial \{Q(t)\}
            }{%
                \partial \mathbf{u}^\dagger(\omega)
            }
        \right|_{f=0}
        =
        \\[20pt]
        \left.
            \dfrac{%
                \partial^2 \{Q(t)\}
            }{%
                \partial \mathbf{u}^\dagger(\omega)
                \partial \mathbf{u}(\omega)
            }
            \dfrac{\partial \mathbf{u}(\omega)}{\partial f(\omega)}
        \right|_{f=0}
            +
        \left.
            \dfrac{%
                \partial^2 \{Q(t)\}
            }{%
                \partial \mathbf{u}^\dagger(\omega)
                \partial f(\omega)
            }
        \right|_{f=0}
    \end{array}
\end{equation}
which constitutes a linear equation for the first-order response of the system
to the perturbation. 
When the frequency $\omega$ is in resonance with an excitation energy of the
system, \cref{eq:linear-response-equation} will result in an infinite
first-order response
\(
    \dfrac{\partial \mathbf{u}(\omega)}{\partial f(\omega)}
\).
From \cref{eq:linear-response-equation}, we find that these poles occur when the
Hessian matrix of the quasi-energy with respect to the wavefunction parameters
\(\mathbf{u}(\omega)\) becomes singular.
We can express this Hessian matrix in the form:
\begin{equation}
    \label{eq:quasi-energy-hessian}
    \left.
        \frac{\partial^2\{Q(t)\}}{%
            \partial \mathbf{u}^\dagger(\omega)
            \partial \mathbf{u}(\omega)
        }
    \right|_{f=0}
    \equiv
    \mathbf{E}
    -
    \omega\,
    \mathbf{M}
\end{equation}
where \(\mathbf{E}\) is the Hessian of the time-averaged electronic energy
\(\{\langle\Psi(t)|\hat{H}|\Psi(t)\rangle\}\) and \(\omega\mathbf{M}\) is the
Hessian of the time-derivative overlap
\(\{\langle\Psi(t)|i\dot{\Psi}(t)\rangle\}\).
The excitation energies of the system $\omega_k$ can therefore be determined by
solving the following generalized eigenvalue equation:
\begin{equation}
    \label{eq:linear-response-energy-eigenvalue-equation}
    \mathbf{E}\mathbf{z}_k
    =
    \omega_k
    \mathbf{M}\mathbf{z}_k
\end{equation}
where \(\mathbf{M}\) serves as the metric matrix.
\cref{eq:linear-response-energy-eigenvalue-equation} allows the determination of
excitation energies for an arbitrary parametrization of $|\Psi(t)\rangle$.

The generalized eigenvectors \(\mathbf{z}_k\) can be used to
compute transition properties for excited states.
In particular, in the exact linear response theory,\cite{Olsen:1985p3235} the transition strength
of the perturbing interaction,
\(
    |\langle\Psi|\hat{V}|\Psi_k\rangle|^2
\),
is equal to the complex residue of the following quantity at
\(\omega\rightarrow\omega_k\):
\begin{equation}
    \langle\!\langle \hat{V}; \hat{V} \rangle\!\rangle_\omega
    \equiv
    \left.
    \mathbf{v}'^\dagger
    \cdot
    \frac{\partial \mathbf{u}(\omega)}{\partial f(\omega)}
    \right|_{f=0}
\end{equation}
This quantity is known as the linear response function and
\(
    \mathbf{v}'
\)
is termed the property gradient vector,\cite{Sauer:2011} which is defined as
follows:
\begin{equation}
    \label{eq:property-gradient-vector}
    \mathbf{v}'
    \equiv
    \left.
    \frac{%
        \partial^2 \{Q(t)\}
    }{%
        \partial \mathbf{u}^\dagger(\omega)
        \partial f(\omega)
    }
    \right|_{f=0}
\end{equation}
Substituting \cref{eq:property-gradient-vector,eq:quasi-energy-hessian} into
\cref{eq:linear-response-equation} and decomposing the quasi-energy Hessian as
\begin{equation}
    \mathbf{E} - \omega\mathbf{M}
    =
    (\mathbf{Z}^\dagger)^{-1}
    (\mathbf{Z}^\dagger \mathbf{M} \mathbf{Z})
    (\boldsymbol\Omega - \omega\mathbf{1})
    (\mathbf{Z})^{-1}
\end{equation}
where \(\mathbf{Z}\) is the matrix of generalized eigenvectors for
\(\mathbf{E}\) and \(\mathbf{M}\) and \(\boldsymbol\Omega\) is the diagonal
matrix of eigenvalues (\cref{eq:linear-response-energy-eigenvalue-equation}), we
obtain the general formula for the transition strengths:
\begin{equation}
    \lim_{\omega\rightarrow \omega_k}
    (\omega-\omega_k)
    \langle\!\langle \hat{V}; \hat{V} \rangle\!\rangle_\omega
    =
    \frac{%
        |\mathbf{z}_k^\dagger \mathbf{v}'|^2
    }{%
        \mathbf{z}_k^\dagger \mathbf{M}\mathbf{z}_k
    }
\end{equation}
In \cref{sec:lr_odc12}, we will use the quasi-energy formalism to derive
equations for the linear-response ODC-12 method (LR-ODC-12).

\subsection{Linear-Response ODC-12}
\label{sec:lr_odc12}

In the ODC-12 method, the time-dependence of the electronic state is specified by the following parameters:
\begin{equation}
    \mathbf{u}(t)
    =
    \begin{pmatrix}
        \mathbf{t}_1(t) \\
        \mathbf{t}_2(t) \\
        \mathbf{t}_1^*(t) \\
        \mathbf{t}_2^*(t)
    \end{pmatrix}
\end{equation}
The ODC-12 electronic Hessian can be written as: 
\begin{equation}
    \label{eq:lr-odc12-hessian-blocks}
    \mathbf{E}
    =
    \begin{pmatrix}
        \mathbf{A}_{11} & \mathbf{A}_{12} & \mathbf{B}_{11} & \mathbf{B}_{12} \\
        \mathbf{A}_{21} & \mathbf{A}_{22} & \mathbf{B}_{21} & \mathbf{B}_{22} \\
        \mathbf{B}_{11}^* & \mathbf{B}_{12}^* & \mathbf{A}_{11}^* & \mathbf{A}_{12}^* \\
        \mathbf{B}_{21}^* & \mathbf{B}_{22}^* & \mathbf{A}_{21}^* & \mathbf{A}_{22}^* \\
    \end{pmatrix}
\end{equation}
where the submatrices are defined in general as
\begin{equation}
    \label{eq:hessian-blocks}
    \mathbf{A}_{nm}
    =
    \left.
    \frac{\partial^2 E}{%
        \partial \mathbf{t}_n^\dagger
        \partial \mathbf{t}_m
    }
    \right|_{f=0}
    ,\ 
    \mathbf{B}_{nm}
    =
    \left.
    \frac{\partial^2 E}{%
        \partial \mathbf{t}_n^\dagger
        \partial \mathbf{t}_m^*
    }
    \right|_{f=0}.
\end{equation}
These complex derivatives relate to the second derivatives of the
electronic energy with respect to variations of the orbitals ($\mathbf{A}_{11}$, $\mathbf{B}_{11}$) and
cumulant parameters ($\mathbf{A}_{22}$, $\mathbf{B}_{22}$).
Similarly, the mixed second derivatives couple variations in the orbitals
and cumulant parameters ($\mathbf{A}_{12}$, $\mathbf{B}_{12}$). 
The metric matrix \(\mathbf{M}\) has a block-diagonal structure, as a
consequence of the linear parametrization of the wavefunction in
\cref{eq:odc12-wavefunction}:
\begin{equation}
    \label{eq:lr-odc12-metric-blocks}
    \mathbf{M}
    =
    \begin{pmatrix}
        \mathbf{S}_{11} & \mathbf{0} & \mathbf{0} & \mathbf{0} \\
        \mathbf{0} & \mathbf{1}_2 & \mathbf{0} & \mathbf{0} \\
        \mathbf{0} & \mathbf{0} & -\mathbf{S}_{11}^* & \mathbf{0} \\
        \mathbf{0} & \mathbf{0} & \mathbf{0} & -\mathbf{1}_2 \\
    \end{pmatrix}
\end{equation}
where
\(
    \mathbf{1}_2
    =
    \langle\Phi|\mathbf{a}_2^\dagger \mathbf{a}_2|\Phi\rangle
\)
is an identity matrix over the space of unique two-body excitations and the
orbital metric is defined as follows:
\begin{equation}
    \label{eq:metric-blocks}
    \omega\mathbf{S}_{11}
    =
    \left.
        \frac{\partial^2 \{\langle\Psi(t)|i\dot\Psi(t)\rangle\}}{%
            \partial \mathbf{t}_1^\dagger(\omega)
            \partial \mathbf{t}_1(\omega)
        }
    \right|_{f=0}
\end{equation}
Equations for all blocks of $\mathbf{E}$, $\mathbf{M}$, and the property
gradient vector $\mathbf{v}'$ are shown explicitly in the Supporting
Information. The computational cost of solving the LR-ODC-12 equations has $\mathcal{O}(O^2V^4)$ scaling (where $O$ and $V$ are the numbers of occupied and virtual orbitals, respectively), which is the same as the computational scaling of the single-state ODC-12 method.
We note that, due to the Hermitian nature of the DCT energy functional
\eqref{eq:energy-expression}, the ODC-12 energy Hessian $\mathbf{E}$ is always
symmetric.
As a result, in the absence of instabilities (i.e., as long as the Hessian is
positive semi-definite), the LR-ODC-12 excitation energies are guaranteed to
have real values. 

To illustrate the derivation of the LR-ODC-12 energy Hessian, let us consider
the diagonal two-body block of \(\mathbf{E}\).
Expressing the energy \eqref{eq:energy-expression} using the cumulant expansion
\eqref{eq:two-body-n-rep} and differentiating with respect to $\mathbf{t}_2$, we
obtain:
\begin{equation}
    \label{eq:odc12-hessian-initial-form}
    \begin{array}{r@{\,}l}
        \mathbf{A}_{22}
        =
        \dfrac{\partial^2 E}{%
            \partial\mathbf{t}_2^\dagger
            \partial\mathbf{t}_2
        }
        =
        &
        f_p^q
        \dfrac{\partial^2 \gamma^p_q}{%
            \partial\mathbf{t}_2^\dagger
            \partial\mathbf{t}_2
        }
        +
        \overline{g}_{pr}^{qs}
        \dfrac{\partial \gamma^p_q}{\partial\mathbf{t}_2^\dagger}
        \dfrac{\partial \gamma^r_s}{\partial\mathbf{t}_2}
        \\[15pt]
        &
        +
        \tfrac{1}{4}
        \overline{g}_{pq}^{rs}
        \dfrac{\partial^2 \lambda^{pq}_{rs}}{%
            \partial\mathbf{t}_2^\dagger
            \partial\mathbf{t}_2
        }
    \end{array}
\end{equation}
where we have introduced the generalized Fock matrix
\(
    f_p^q
    \equiv
    h_p^q
    +
    \overline{g}_{pr}^{qs}
    \gamma^r_s
\).
The derivatives of the one-body density matrix can be expressed in terms of the derivatives of the density cumulant
\begin{equation}
    \begin{array}{r@{\,}l}
    \label{eq:a22_odc12}
	\mathbf{A}_{22}
        =
        &
        \mathcal{F}_p^q
        \dfrac{\partial^2 \lambda^{pt}_{qt}}{%
            \partial\mathbf{t}_2^\dagger
            \partial\mathbf{t}_2
        }
        +
        \mathcal{G}_{pr}^{qs}
        \dfrac{\partial \lambda^{pt}_{qt}}{\partial\mathbf{t}_2^\dagger}
        \dfrac{\partial \lambda^{ru}_{su}}{\partial\mathbf{t}_2}
        \\[15pt]
        &
        +
        \tfrac{1}{4}
        \overline{g}_{pq}^{rs}
        \dfrac{\partial^2 \lambda^{pq}_{rs}}{%
            \partial\mathbf{t}_2^\dagger
            \partial\mathbf{t}_2
        }
    \end{array}
\end{equation}
where the intermediates $\mathcal{F}_p^q$ and $\mathcal{G}_{pr}^{qs}$ can be computed using a transformation of the one- and two-electron integrals to the natural spin-orbital basis (see \cref{sec:appendix} for details).
These cumulant derivatives are straightforward to evaluate from \cref{eq:odc12-wavefunction,eq:cumulant-parametrization} using either algebraic or diagrammatic techniques. 

Next, we outline the derivation of the metric $\mathbf{M}$ (see Supporting Information for more details).
For the one-electron block of the metric,
substituting \cref{eq:odc12-wavefunction} into \cref{eq:metric-blocks} gives
\begin{equation}
    \label{eq:odc12-s11-metric}
    \begin{array}{r@{\,}l}
        \omega
        \mathbf{S}_{11}
        =
        &
        \left.
            \dfrac{1}{2}
            \dfrac{%
                \partial^2
                \{\langle\Psi|
                    [\hat{T}_1^\dagger(t), i\hat{\dot{T}}_1(t)]
                |\Psi\rangle\}
            }{%
                \partial \mathbf{t}_1^\dagger(\omega)
                \partial \mathbf{t}_1(\omega)
            }
        \right|_{f=0}
        \\[20pt]
        &
        -
        \left.
            \dfrac{1}{2}
            \dfrac{%
                \partial^2
                \{\langle\Psi|
                    [i\hat{\dot{T}}_1^\dagger(t), \hat{T}_1(t)]
                |\Psi\rangle\}
            }{%
                \partial \mathbf{t}_1^\dagger(\omega)
                \partial \mathbf{t}_1(\omega)
            }
        \right|_{f=0}
    \end{array}
\end{equation}
where we have assumed that we are working in the variational orbital basis so that
\(
    \hat{T}_1(t)|_{f=0}
    =
    0
\),
and
\(
    \Psi
    =
    \Psi(t)|_{f=0}
\)
denotes the ground state wavefunction.
Using the Fourier expansion of the $\mathbf{t}_1(t)$ parameters
(\cref{eq:parameter-fourier-expansion}), the gradients of the time derivatives
can be evaluated as:
\begin{equation}
    \label{eq:time_deriv_1}
    \left.
    \frac{%
        \partial i \hat{\dot{T}}_1(t)
    }{%
        \partial \mathbf{t}_1(\omega)
    }
    \right|_{f=0}
    =
    +
    \omega
    \mathbf{a}_1
    e^{-i\omega t}
\end{equation}
\begin{equation}
    \label{eq:time_deriv_2}
    \left.
    \frac{%
        \partial i \hat{\dot{T}}_1^\dagger(t)
    }{%
        \partial \mathbf{t}_1^\dagger(\omega)
    }
    \right|_{f=0}
    =
    -
    \omega
    \mathbf{a}_1^\dagger
    e^{+i\omega t}
\end{equation}
Substituting \cref{eq:time_deriv_1,eq:time_deriv_2} into \cref{eq:odc12-s11-metric} and evaluating the gradients
of \(\hat{T}_1\) and \(\hat{T}_1^\dagger\) similarly gives the final working equation for
the one-body metric:
\begin{equation}
    \begin{array}{r@{\,}l}
        \omega
        (\mathbf{S}_{11})_{ia,jb}
        &=
        \omega
        \langle\Psi|
        [a_a^i, a_j^b]
        |\Psi\rangle
        \\[10pt]
        &=
        \omega
        (
            \delta^b_a
            \gamma^i_j
            -
            \delta^i_j
            \gamma^b_a
        )
    \end{array}
\end{equation}
The metric contributions involving the second derivatives with respect to \(\mathbf{t}_2\) have been
determined using the linearized doubles parametrization of the wavefunction in \Cref{eq:odc12-wavefunction}. Since the ODC-12 energy is correct to the third order in perturbation theory\cite{Sokolov:2014p074111}, these \(\mathbf{t}_2\) contributions to the metric are also truncated at the third order. Using this approximation, we find that in LR-ODC-12 the \(\mathbf{t}_2\) second derivative contributions to the metric vanish. These results are in agreement with the expressions for the metric matrix elements in time-dependent unitary coupled-cluster doubles theory,\cite{Walz:2012p052519} which do not contain \(\mathbf{t}_2\) contributions up to the third order in perturbation theory. The mixed \(\mathbf{t}_1\)-\(\mathbf{t}_2\) (orbital-cumulant) blocks of the metric matrix are zero at any order of perturbation theory.

\subsection{Linear-Response OLCCD}
\label{sec:olccd}
As we discussed in \cref{sec:dct}, the orbital-optimized linearized coupled cluster doubles method (OLCCD) can be considered as an approximation to the ODC-12 method where all of the non-linear $\mathbf{t}_2$ terms are neglected in the stationarity conditions. Similarly, we can formulate the linear-response OLCCD method (LR-OLCCD) by linearizing the LR-ODC-12 equations. This simplifies the expressions for the electronic Hessian blocks that involve the second derivatives with respect to $\mathbf{t}_2$. For example, for the $\mathbf{A}_{22}$ block, we obtain:
\begin{equation}
    \label{eq:a22_olccd}
    \mathbf{A}_{22}
    =
    (f_0)_i^j
    \dfrac{\partial^2 \lambda^{ir}_{jr}}{%
        \partial\mathbf{t}_2^\dagger
        \partial\mathbf{t}_2
    }
    -
    (f_0)_a^b
    \dfrac{\partial^2 \lambda^{ar}_{br}}{%
        \partial\mathbf{t}_2^\dagger
        \partial\mathbf{t}_2
    }
    +
    \tfrac{1}{4}
    \overline{g}_{pq}^{rs}
    \dfrac{\partial^2 \lambda^{pq}_{rs}}{%
        \partial\mathbf{t}_2^\dagger
        \partial\mathbf{t}_2
    }
\end{equation}
where
\(
    (f_0)_p^q
    =
    h_p^q
    +
    \overline{g}_{pi}^{qi}
\)
is the usual (mean-field) Fock operator.
Comparing \cref{eq:a22_olccd} with \cref{eq:a22_odc12} from the LR-ODC-12
method, we observe that the former equation can be obtained from the latter by
replacing the $\mathcal{F}_p^q$ intermediates with the mean-field Fock matrix
elements and ignoring the term that depends on $\mathcal{G}_{pr}^{qs}$.
These simplifications arise from the fact that the $\mathcal{F}_p^q$ and
$\mathcal{G}_{pr}^{qs}$ intermediates contain high-order $\mathbf{t}_2$
contributions that are not included in the linearized LR-OLCCD formulation (see
\cref{sec:appendix} and Ref.\@ \citenum{Sokolov:2013p024107} for details).
For the $\mathbf{B}_{22}$ block, we find that all of the Hessian elements are
zero.
A complete set of working equations for LR-OLCCD is given in the Supporting
Information.

\section{Computational Details}
\label{sec:comp_details}
The LR-ODC-12 and LR-OLCCD methods were implemented as a standalone Python
program, which was interfaced with \textsc{Psi4}\cite{Parrish:2017p3185} and
\textsc{Pyscf}\cite{Sun:2018pe1340} to obtain the one- and two-electron
integrals.
To compute excitation energies, our implementation utilizes the multi-root
Davidson algorithm,\cite{Davidson:1975p87,Liu:1978p49} which solves the
generalized eigenvalue problem
\eqref{eq:linear-response-energy-eigenvalue-equation} by progressively growing
an expansion space for the \(n_\mathrm{root}\) lowest generalized eigenvectors
of the electronic Hessian and the metric matrix.
A key feature of this algorithm is that it avoids storing the Hessian and metric
matrices, significantly reducing the amount of memory required by the
computations.
Our implementation of the energy Hessian was validated by computing the static
response function for a dipole perturbation (i.e., the dipole polarizability):
\begin{equation}
\label{eq:dip_polariz_analytic}
    \langle\!\langle\hat{V}; \hat{V}\rangle\!\rangle_0
    =
    -
    \mathbf{v}'^\dagger
    \mathbf{E}^{-1}
    \mathbf{v}'
\end{equation}
This quantity can be evaluated numerically as a derivative of the ground state energy
\begin{equation}
\label{eq:dip_polariz_numerical}
    \langle\!\langle\hat{V}; \hat{V}\rangle\!\rangle_0
    =
    \left.
    \frac{%
        d^2 E
    }{%
        df^2
    }
    \right|_{f=0}
\end{equation}
by perturbing the one-electron integrals
\(
    h_p^q
    \leftarrow
    h_p^q
    +
    f
    v_p^q
\)
with the integrals of the perturbing dipole operator (\(v_p^q\)), and solving the
ODC-12 (or OLCCD) equations for different values of \(f\).
For the dipole polarizability of the water molecule along its \(C_2\) symmetry
axis, the values of $\langle\!\langle\hat{V}; \hat{V}\rangle\!\rangle_0$
computed using \cref{eq:dip_polariz_analytic,eq:dip_polariz_numerical} matched
to \(10^{-9}\ \au\)

We used \textsc{Q-Chem} 4.4\cite{qchem:44} to obtain results from
equation-of-motion coupled cluster theory with single and double excitations
(EOM-CCSD) and EOM-CCSD with triple excitations in the EOM part [EOM-CC(2,3)].
The \textsc{MRCC} program\cite{MRCC} was used to obtain results for
equation-of-motion coupled cluster theory with up to full triple excitations
(EOM-CCSDT). All electrons were correlated in all computations. We used tight
convergence parameters in all ground-state ($10^{-8}$ \eh) and excited-state
computations ($10^{-5}$ \eh). In \cref{sec:two_electron,sec:vert_excit}, the
augmented aug-cc-pVTZ and d-aug-cc-pVTZ basis sets of Dunning and co-workers
were employed.\cite{Kendall:1992p6796} For alkenes (\cref{sec:alkenes}), the
ANO-L-pVXZ (X = D, T) basis sets\cite{Widmark:1990p291} were used as in Ref.
\citenum{Daday:2012p4441}. To compute vertical excitation energies in
\cref{sec:vert_excit}, geometries of molecules were optimized using ODC-12 (for
LR-ODC-12), OLCCD (for LR-OLCCD), or CCSD [for EOM-CCSD, EOM-CC(2,3), and
EOM-CCSDT].
For the alkenes in \cref{sec:alkenes}, frozen-core MP2/cc-pVQZ geometries were
used as in Refs.\citenum{Daday:2012p4441} and \citenum{Zimmerman:2017p4712}.

\section{Results}
\label{sec:results}

\subsection{Size-Intensivity of the LR-ODC-12 Energies}
\label{sec:size_intensivity}

\begin{table*}[h!]
    \caption{Ground-state energies (in \eh) and vertical excitation energies (in
    eV) for the four lowest-energy excited states of the CO molecule and
noninteracting systems of CO with Ne atoms (\mbox{CO $+$ $n$Ne}, $n$ = 1, 2, 3)
computed using the ODC-12 and LR-ODC-12 methods (cc-pVDZ basis set). Also shown results for two noninteracting CO molecules (CO + CO). The noninteracting systems were separated from each other by \mbox{10000 \AA} and
the C--O bond distance was set to 1.12547 \AA\@. Results demonstrate size-intensivity of the LR-ODC-12 excitation energies. }
   \label{tab:size_int}
    \begin{tabular}{cccccc}
        \hline
        \hline
        & CO & CO $+$ Ne & CO $+$ 2Ne & CO $+$ 3Ne & CO $+$ CO \\
        \hline
        \({X}\,^1\Sigma^+\) & $-$113.051282 & $-$241.730913 & $-$370.410543 & $-$499.090174 &$-$226.102565\\
        \({}^3\Pi\) & 6.48597 & 6.48597 & 6.48597 & 6.48597 &6.48597\\
        \({}^3\Sigma^+\) & 8.41225 & 8.41225 & 8.41225 & 8.41225 &8.41225\\
        \({}^1\Pi\) & 8.90866 & 8.90866 & 8.90866 & 8.90866 &8.90866\\
        \({}^3\Delta\) & 9.33189 & 9.33189 & 9.33189 & 9.33189 &9.33189\\
        \hline
        \hline
    \end{tabular}
\end{table*}

In \cref{sec:dct}, we mentioned that all DCT methods are by construction {\it
size-extensive}, meaning that their electronic energies scale linearly with the
number of electrons.
In this section, we demonstrate that the LR-ODC-12 excitation energies are {\it size-intensive}, i.e.\@ they satisfy the following property: $E(A^*+B) = E(A^*) + E(B)$, where $A$ and $B$ are two noninteracting fragments in their corresponding ground states and $A^*$ is the fragment $A$ in an excited state. \cref{tab:size_int} shows the ODC-12 ground-state energies and the LR-ODC-12 excitation energies for the CO molecule and noninteracting systems composed of CO and the neon atoms separated by 10000 \AA\ (\mbox{CO $+$ $n$Ne}, $n$ = 1, 2, 3), as well as for two noninteracting CO molecules (CO + CO). The scaling of the ODC-12 energies with the number of electrons for the ground \({X}\,^1\Sigma^+\) electronic state is perfectly linear up to $10^{-8}$ \eh, which is the convergence parameter used in our ODC-12 computations. Upon the addition of the noninteracting atoms and molecules, the excitation energies of the CO molecule remain constant up to the convergence threshold set in LR-ODC-12 ($10^{-6}$  eV). These results provide numerical evidence that the LR-ODC-12 excitation energies are size-intensive.

\subsection{H$_2$ Dissociation}
\label{sec:two_electron}

\begin{figure*}[h!]
   \label{fig:H2_diss}
   \subfloat[]{\label{fig:H2_diss_odc}\includegraphics[width=0.45\textwidth]{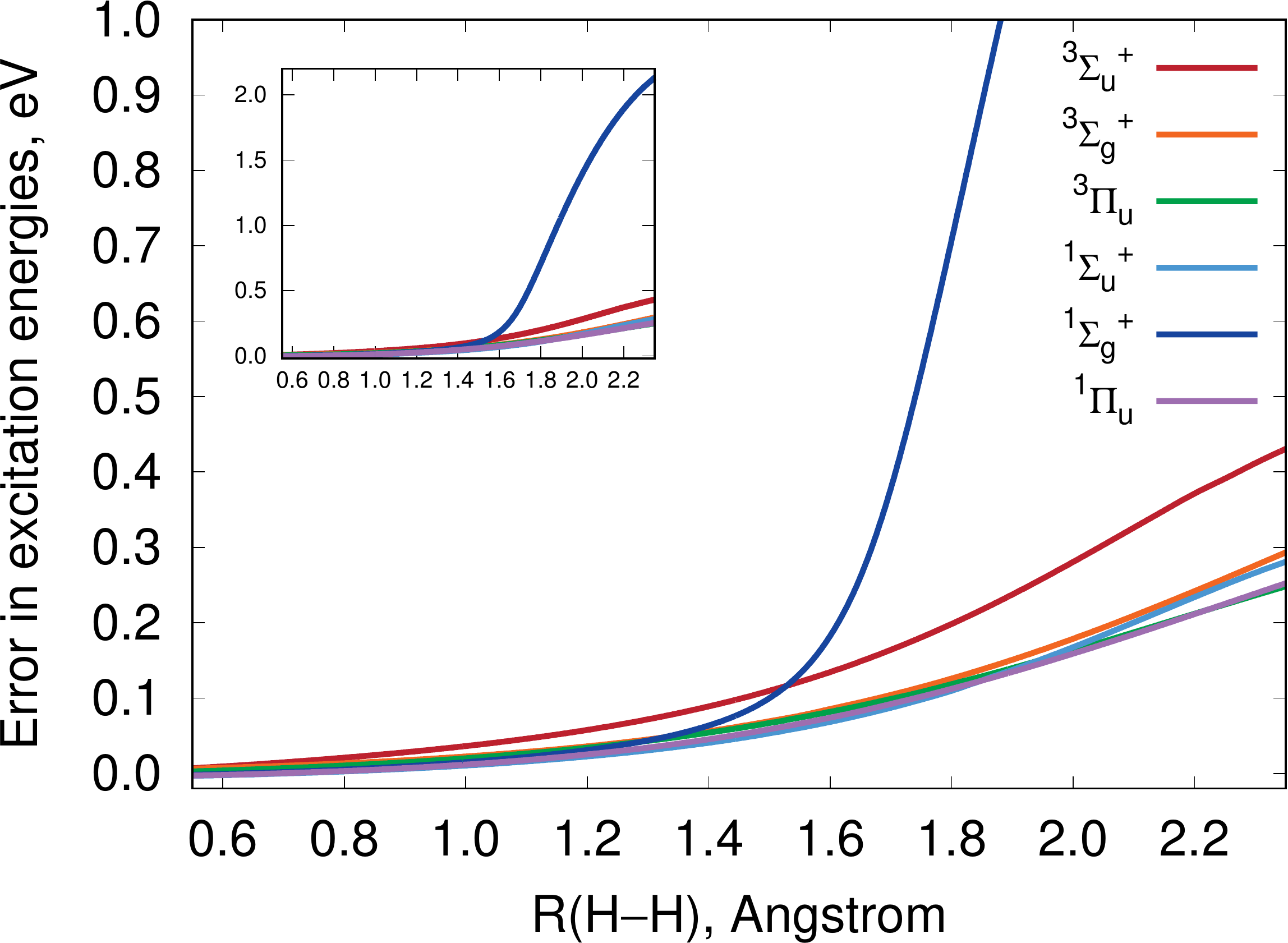}} \qquad
   \subfloat[]{\label{fig:H2_diss_olccd}\includegraphics[width=0.45\textwidth]{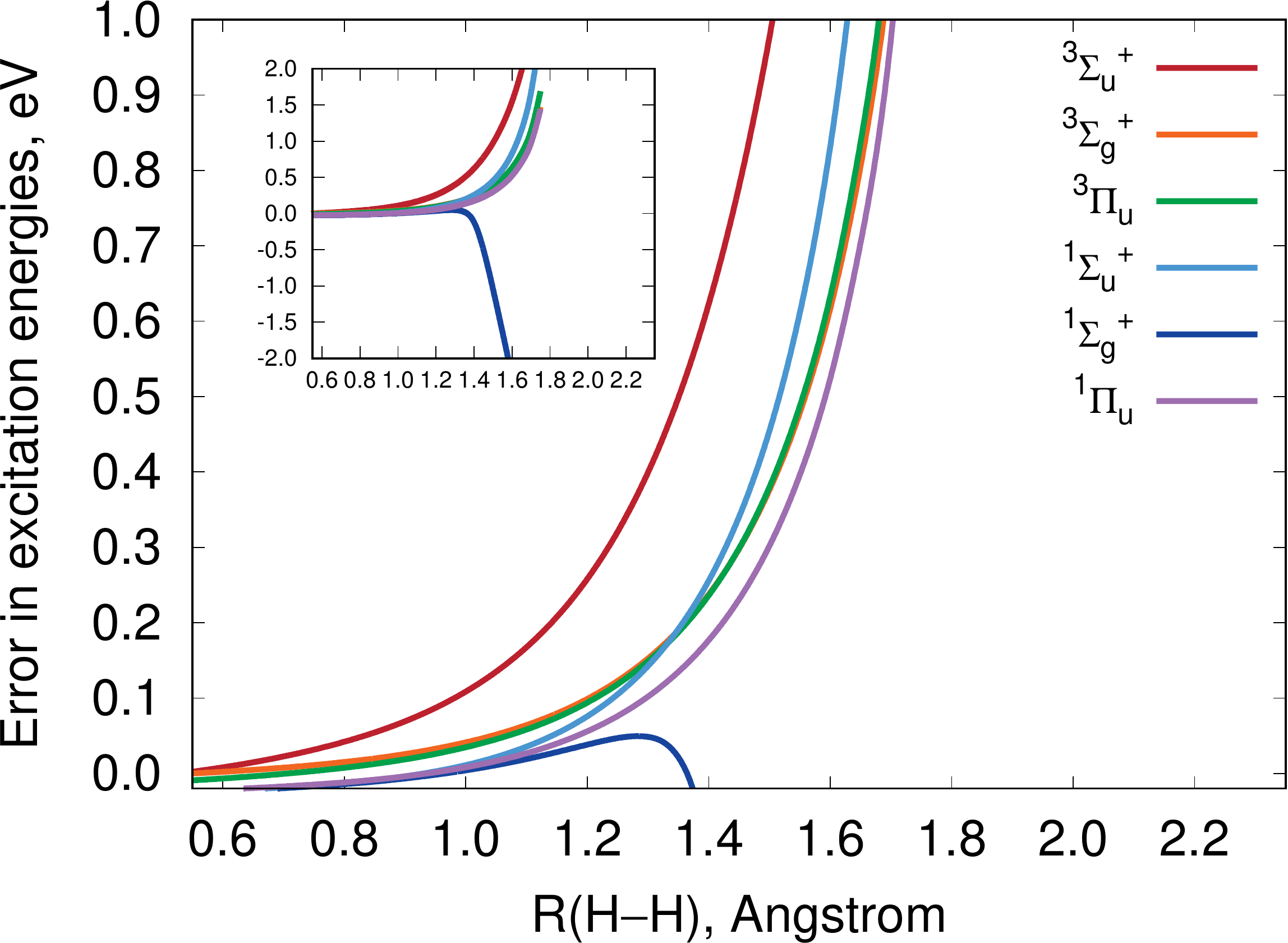}} \quad
   \captionsetup{justification=raggedright,singlelinecheck=false}
   \caption{Errors in vertical excitation energies (eV) for six lowest-lying electronic states of \ce{H2} computed using LR-ODC-12 (\ref{fig:H2_diss_odc}) and LR-OLCCD (\ref{fig:H2_diss_olccd}) as a function of the H--H bond length, relative to full configuration interaction. All methods employed the d-aug-cc-pvtz basis set. In each figure, the inset shows the same plot for a larger range of errors.}
   \label{fig:algorithm_errors}
\end{figure*}

One of the desirable properties of an electronic structure method is exactness
for two-electron systems. While the ODC-12 method is not exact for two-electron
systems, it has been shown to provide a very good description of the
ground-state \ce{H2} dissociation curve, with errors of $\sim$ 1 \kcal with
respect to full configuration interaction (FCI) near the dissociation
limit.\cite{Sokolov:2013p204110} Here, we investigate the performance of
LR-ODC-12 for the excited states of \ce{H2}. \cref{fig:H2_diss_odc} shows the
errors in vertical excitation energies for six lowest-lying electronic states as
a function of the \ce{H-H} distance, relative to FCI\@. The FCI energies were
computed using the EOM-CCSD method, which is exact for two-electron systems. At
the equilibrium geometry ($r_e$ = 0.742 \AA) the errors in excitation energies
for all states do not exceed 0.02 eV. Between 0.6 and 1.45 \AA\
(\mbox{$r\approx2r_e$}), the LR-ODC-12 excitation energies remain in
good agreement with FCI, with errors less than 0.1 eV for all states. In this
range, the largest error is observed for the $^3\Sigma_u^+$ state. For $r$ $\ge$
1.5 \AA, the error in the $^1\Sigma_g^+$ excited state energy rapidly increases
from 0.10 eV (at 1.5 \AA) to 2.13 eV (at 2.35 \AA), while for other states the
errors increase much more slowly. Analysis of the FCI wavefunction for the
$^1\Sigma_g^+$ state shows a significant contribution from the
$(1\sigma_g)^2\rightarrow(1\sigma_u)^2$ double excitation already at $r$ $=$
1.55 \AA\@. This contribution becomes dominant for $r$ $\ge$ 1.75 \AA\@. Thus, the large LR-ODC-12 errors observed for the $^1\Sigma_g^+$ state are likely due to the increasingly large double-excitation character of this electronic state at long \ce{H-H} bond distances. The second largest error near the dissociation is observed for the $^3\Sigma_u^+$ state (0.43 eV). For other electronic states, smaller errors of $\sim$ 0.25 eV are observed near the dissociation.

The importance of the non-linear terms in the LR-ODC-12 equations can be
investigated by comparing the LR-ODC-12 and LR-OLCCD results.
\cref{fig:H2_diss_olccd} shows the errors in the LR-OLCCD vertical excitation
energies as a function of the \ce{H-H} bond length. Although near the
equilibrium geometry the performance of LR-OLCCD and LR-ODC-12 is similar, the
LR-OLCCD errors increase much faster with increasing \ce{H-H} distance compared
to LR-ODC-12. At $r$ = 1.3 \AA, the LR-OLCCD error for the $^3\Sigma_u^+$ state
(0.4 eV) is almost six times larger than the corresponding error from LR-ODC-12
(0.07 eV). For $r$ $\ge$ 1.35 \AA, the LR-OLCCD errors for all excitation
energies show very steep increase in magnitude, ranging from 1.5 to 4.7 eV
already at $r$ = 1.75 \AA\@. We were unable to converge the LR-OLCCD equations
for $r$ $\ge$ 1.80 \AA\@.
Overall, our results demonstrate that the non-linear terms in LR-ODC-12
significantly improve the description of the excited states at long \ce{H-H}
distances where the electron correlation effects are stronger.

\subsection{Benchmark: Small Molecules}
\label{sec:vert_excit}

\begin{table*}[h!]
    \caption{Errors in vertical excitation energies (eV) for singlet states computed
        using LR-OLCCD, LR-ODC-12, and EOM-CCSD, relative to EOM-CCSDT
        (aug-cc-pVTZ basis set).
        All electrons were correlated in all computations.
        Also shown are mean absolute errors (\mae), standard deviations
        (\std), and maximum absolute errors (\maxe) computed for each method.}
   \label{tab:vertical_singlet}
\begin{threeparttable}
    \begin{tabular}{clcccc}
        \hline
        \hline
        && \(\Delta\)EOM-CCSD & \(\Delta\)LR-OLCCD & \(\Delta\)LR-ODC-12 & 
        EOM-CCSDT \\
        \hline
        \ce{N2}
        & \({}^1\Pi_g\)        & \( \)0.18 & \( \)0.08 & \( \)0.20 &  9.29 \\
        & \({}^1\Sigma_u^-\)   & \( \)0.23 & \( \)0.15 & \( \)0.09 &  9.84 \\
        & \({}^1\Delta_u\)     & \( \)0.26 & \( \)0.14 & \( \)0.10 & 10.26 \\
        \ce{CO}                            
        & \({}^1\Pi\)          & \( \)0.16 & \( \)0.09 & \( \)0.17 &  8.46 \\
        & \({}^1\Sigma^-\)     & \( \)0.19 & \(-\)0.10 & \(-\)0.01 &  9.89 \\
        & \({}^1\Delta\)       & \( \)0.19 & \(-\)0.22 & \(-\)0.05 & 10.03 \\
        \ce{HCN}                           
        & \({}^1\Sigma^-\)     & \( \)0.16 & \( \)0.05 & \( \)0.00 & 8.25 \\
        & \({}^1\Delta\)       & \( \)0.17 & \( \)0.04 & \( \)0.01 & 8.61 \\
        & \({}^1\Pi\)          & \( \)0.17 & \( \)0.05 & \( \)0.20 & 9.12 \\
        \ce{HNC}                           
        & \({}^1\Pi\)          & \( \)0.15 & \(-\)0.01 & \( \)0.10 & 8.13 \\
        & \({}^1\Sigma^+\)     & \( \)0.24 & \( \)0.05 & \( \)0.12 & 8.46 \\
        & \({}^1\Sigma^-\)     & \( \)0.15 & \(-\)0.09 & \( \)0.04 & 8.67 \\
        & \({}^1\Delta\)       & \( \)0.15 & \(-\)0.18 & \(-\)0.03 & 8.84 \\
        \ce{C2H2}                          
        & \({}^1\Sigma_u^-\)   & \( \)0.12 & \( \)0.06 & \( \)0.02 & 7.11 \\
        & \({}^1\Delta_u\)     & \( \)0.10 & \( \)0.07 & \( \)0.03 & 7.45 \\
        \ce{H2CO}                                                 
        & \({}^1\mathrm{A_2}\) & \( \)0.10 & \(-\)0.07 & \( \)0.02 & 3.95 \\
        \hline
        \mae & &0.17  &0.09 &0.08   &  \\		
        \std   & &0.05  & 0.11 &0.08  &  \\ 		
        \maxe & &0.26  & 0.22 &0.20  &  \\ 		
        \hline
        \hline
    \end{tabular}
\end{threeparttable}
\end{table*}

\begin{table*}[h!]
       \caption{Errors in vertical excitation energies (eV) for triplet states computed
        using LR-OLCCD, LR-ODC-12, and EOM-CCSD, relative to EOM-CCSDT
        (aug-cc-pVTZ basis set).
        All electrons were correlated in all computations.
        Also shown are mean absolute errors (\mae), standard deviations
        (\std), and maximum absolute errors (\maxe) computed for each method.}
   \label{tab:vertical_triplet}
\begin{threeparttable}
    \begin{tabular}{clcccc}
        \hline
        \hline
        && \(\Delta\)EOM-CCSD & \(\Delta\)LR-OLCCD & \(\Delta\)LR-ODC-12 &
        EOM-CCSDT \\
        \hline
        \ce{N2}
        & \({}^3\Sigma_u^+\)   & \( \)0.11 & \( \)0.04 & \(-\)0.02 &  7.63 \\
        & \({}^3\Pi_g\)        & \( \)0.15 & \( \)0.06 & \( \)0.11 &  8.00 \\
        & \({}^3\Delta_u\)     & \( \)0.17 & \( \)0.08 & \( \)0.03 &  8.82 \\
        & \({}^3\Sigma_u^-\)   & \( \)0.28 & \( \)0.03 & \( \)0.01 &  9.63 \\
        & \({}^3\Pi_u\)        & \( \)0.14 & \(-\)0.01 & \( \)0.10 & 11.18 \\
        \ce{CO}                            
        & \({}^3\Pi\)          & \( \)0.12 & \( \)0.06 & \( \)0.08 &  6.27 \\
        & \({}^3\Sigma^+\)     & \( \)0.05 & \(-\)0.03 & \(-\)0.03 &  8.38 \\
        & \({}^3\Delta\)       & \( \)0.11 & \(-\)0.07 & \(-\)0.03 &  9.21 \\
        & \({}^3\Sigma^-\)     & \( \)0.19 & \(-\)0.18 & \(-\)0.06 &  9.72\tnote{a} \\
        \ce{HCN}                           
        & \({}^3\Sigma^+\)     & \( \)0.05 & \(-\)0.04 & \(-\)0.10 &  6.40 \\
        & \({}^3\Delta\)       & \( \)0.13 & \(-\)0.02 & \(-\)0.06 &  7.40 \\
        & \({}^3\Pi\)          & \( \)0.10 & \( \)0.08 & \( \)0.06 &  8.01 \\
        & \({}^3\Sigma^-\)     & \( \)0.16 & \(-\)0.10 & \(-\)0.05 &  8.15\tnote{a} \\
        \ce{HNC}                                                      
        & \({}^3\Pi\)          & \( \)0.09 & \( \)0.00 & \( \)0.03 &  6.06 \\
        & \({}^3\Sigma^+\)     & \( \)0.04 & \(-\)0.09 & \(-\)0.11 &  7.20 \\
        & \({}^3\Delta\)       & \( \)0.10 & \(-\)0.14 & \(-\)0.11 &  8.02 \\
        & \({}^3\Sigma^+\)     & \( \)0.22 & \(-\)0.05 & \( \)0.04 &  8.38 \\
        & \({}^3\Sigma^-\)     & \( \)0.15 & \(-\)0.02 & \( \)0.11 &  8.56\tnote{a} \\
        \ce{C2H2}                                                     
        & \({}^3\Sigma_u^+\)   & \( \)0.01 & \(-\)0.02 & \(-\)0.08 &  5.52 \\
        & \({}^3\Delta_u\)     & \( \)0.08 & \(-\)0.02 & \(-\)0.05 &  6.41 \\
        & \({}^3\Sigma_u^-\)   & \( \)0.10 & \(-\)0.03 & \(-\)0.05 &  7.10\tnote{a} \\
        \ce{H2CO}                                                     
        & \({}^3\mathrm{A_2}\) & \( \)0.04 & \(-\)0.02 & \( \)0.01 &  3.56 \\
        & \({}^3\mathrm{A_1}\) & \( \)0.02 & \(-\)0.06 & \(-\)0.14 &  6.06 \\
        \hline
        \mae & & 0.11 &0.05  &0.06  &  \\ 		
        \std   & & 0.06 &0.07  &0.07  &  \\ 		
        \maxe & &0.28  & 0.18 &0.14  &  \\ 		
        \hline
        \hline
    \end{tabular}
    \begin{tablenotes}
    \item[a] For CO, HCN, HNC, and \ce{C2H2}, the ${}^3\Sigma^-$ (${}^3\Sigma_u^-$) excitation energies were obtained from EOM-CC(2,3), which energies were shifted to reproduce the EOM-CCSDT energy for the ${}^1\Sigma^-$ (${}^1\Sigma_u^-$) state.
    \end{tablenotes}    
\end{threeparttable}
\end{table*}

\begin{figure}
    \caption{%
        Mean absolute deviations (\mae) and standard deviations from the mean
        signed error (\std) for vertical excitation energies 
        (\cref{tab:vertical_singlet,tab:vertical_triplet}) computed using LR-OLCCD, LR-ODC-12,
        and EOM-CCSD, relative to EOM-CCSDT (aug-cc-pVTZ basis set).
        The \mae value is represented as a height of each colored box, while the \std value is depicted as a radius of the black vertical bar. 
    }
    \label{fig:vertical_mae}
    \includegraphics[width=8.5cm]{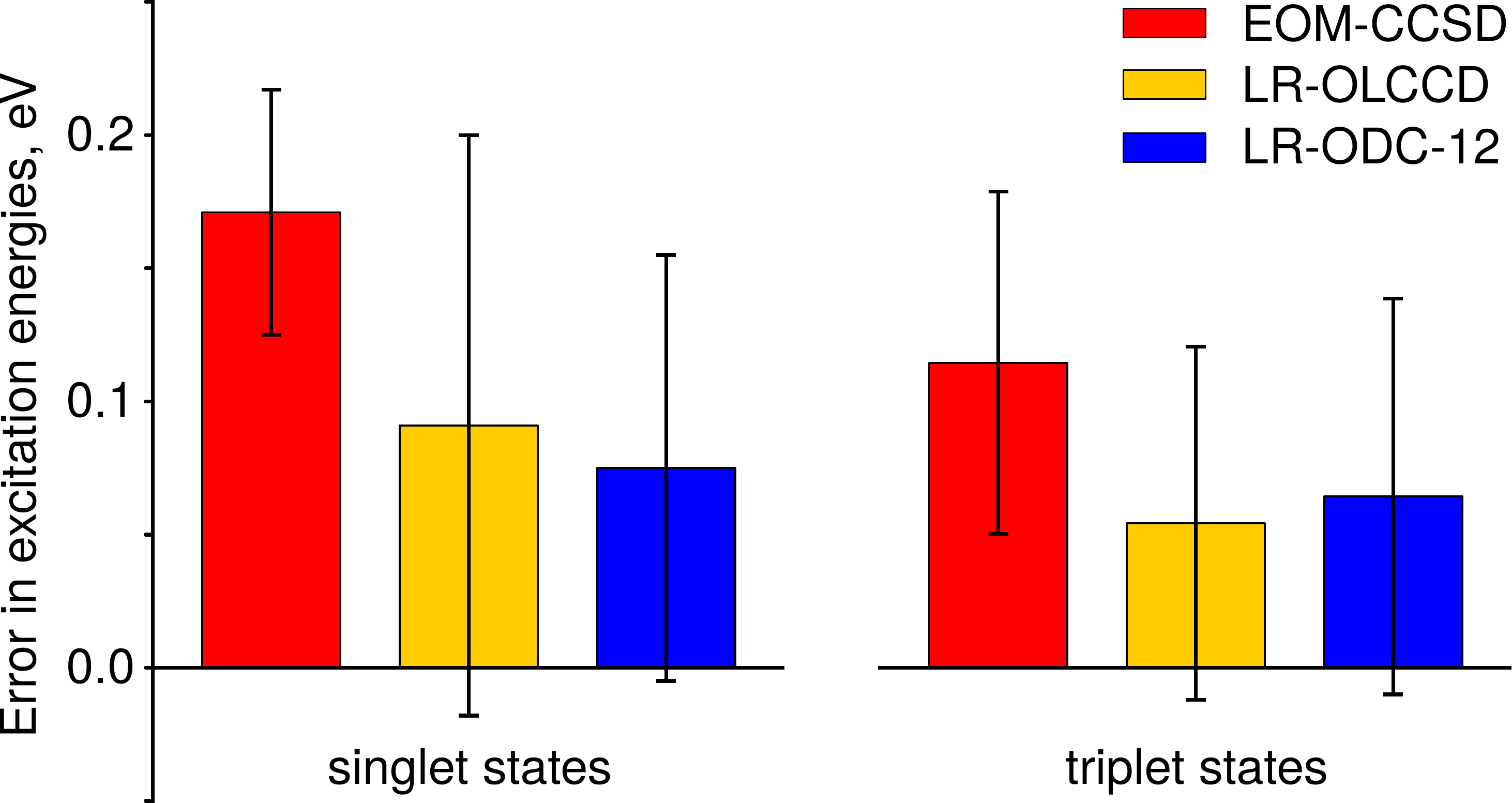}
\end{figure}

Here, we benchmark the performance of LR-ODC-12 for vertical excitation
energies in several small molecules: \ce{N2}, \ce{CO}, \ce{HCN}, \ce{HNC},
\ce{C2H2}, and \ce{H2CO}. 
\cref{tab:vertical_singlet,tab:vertical_triplet} show the errors in excitation
energies computed using EOM-CCSD, LR-OLCCD, and LR-ODC-12 for the singlet and
triplet excited states, respectively, relative to the results from EOM-CCSDT\@. To measure the performance of each method, we computed the mean absolute errors (\mae) and the standard deviations from the average signed error (\std), shown in \cref{fig:vertical_mae}.

For the singlet electronic states (\cref{tab:vertical_singlet}), the excitation energies computed using LR-ODC-12 are in better agreement with EOM-CCSDT than those obtained from EOM-CCSD, on average. This is evidenced by \mae, which is smaller for LR-ODC-12 compared to EOM-CCSD by a factor of two (\mae = 0.08 and 0.17 eV, respectively). The LR-ODC-12 errors exceed 0.10~eV for only four states, with a maximum error \maxe = 0.20~eV. EOM-CCSD has a minimum error of 0.10 eV, shows errors greater than 0.10~eV for 14 states, and has \maxe = 0.26~eV. EOM-CCSD shows a somewhat smaller \std compared to that of LR-ODC-12 (\std = 0.05 and 0.08 eV, respectively).

For the triplet states (\cref{tab:vertical_triplet}), LR-ODC-12 is again superior to EOM-CCSD, on average, with \mae = 0.06 and 0.11 eV for the two methods, respectively. LR-ODC-12 has errors larger than 0.10~eV for five states with \maxe = 0.14 eV, whereas EOM-CCSD exceeds 0.10 eV error for 12 states and shows \maxe = 0.28 eV. For linear molecules, EOM-CCSD exhibits consistently poor results for the \({}^3\Sigma^-\) electronic states, while the performance of LR-ODC-12 for different electronic states is similar. Notably, all EOM-CCSD excitation energies overestimate the EOM-CCSDT values, while the LR-ODC-12 energies are centered around the reference energies, suggesting that LR-ODC-12 provides a more balanced description of the ground and excited states.

Comparing LR-ODC-12 with LR-OLCCD, we see that both methods show very similar results for the triplet states (\mae = 0.06 and 0.05 eV, respectively), with noticeable differences observed only for the \({}^3\Sigma^-\) states. For the singlet electronic states, LR-OLCCD shows a somewhat larger \mae = 0.09 eV and \std = 0.11 eV compared to LR-ODC-12 (\mae = 0.08 eV and \std = 0.08 eV). In this case, significant differences are observed for the \({}^1\Pi\) states of \ce{N2} and HCN, \({}^1\Sigma^-\) of HNC, and \({}^1\Delta\) of CO and HNC, indicating that the non-linear terms included in LR-ODC-12 are important for these electronic states. 

\subsection{Ethylene, Butadiene, and Hexatriene}
\label{sec:alkenes}

\begin{table*}[h!]
       \caption{Ground-state total energies (\eh) and vertical excitation energies (eV) computed using LR-OLCCD, LR-ODC-12, and EOM-CCSD for the low-lying electronic states of ethylene (\ce{C2H4}), butadiene (\ce{C4H6}), and hexatriene (\ce{C6H8}). Computations employed the ANO-L-pVDZ (for \ce{C4H6} and \ce{C6H8}) and ANO-L-pVTZ (for \ce{C2H4}) basis sets and the MP2/cc-pVQZ optimized geometries. For LR-OLCCD and LR-ODC-12, oscillator strengths of the allowed transitions are given in parentheses. All electrons were correlated in all computations.}
   \label{tab:alkenes}
\begin{threeparttable}
    \begin{tabular}{clccccc}
        \hline
        \hline
        && EOM-CCSD & LR-OLCCD & LR-ODC-12 & SHCI\tnote{a} \\
        \hline
        \ce{C2H4}
        & \(1{}^1\mathrm{A_{1g}}\) & $-$78.441518 &  $-$78.455352     &     $-$78.450874     & $-$78.4381  \\ 
        & \(1{}^3\mathrm{B_{1u}}\) & 4.46 & 4.66         & 4.52         & 4.59  \\ 
        & \(1{}^1\mathrm{B_{1u}}\) & 8.14 & 8.20 (1.8) & 8.13 (1.9) & 8.05 \\
        \hline                           
        \ce{C4H6}                        
        & \(1{}^1\mathrm{A_{g}}\)  & $-$155.546920 &  $-$155.568356        &  $-$155.559882  & $-$155.5582 \\
        & \(1{}^3\mathrm{B_{u}}\)  & 3.20 & 3.58         & 3.43         & 3.37 \\
        & \(1{}^1\mathrm{B_{u}}\)  & 6.53 & 6.76 (4.2) & 6.67 (4.4) & 6.45 \\
        & \(2{}^1\mathrm{A_{g}}\)  & 7.28 & 7.14         & 6.81         & 6.58 \\
        \hline                           
        \ce{C6H8}                        
        & \(1{}^1\mathrm{A_{g}}\)  & $-$232.737880 &  $-$232.771738        &  $-$232.758675        &  $-$232.7567\\
        & \(1{}^3\mathrm{B_{u}}\)  & 2.64 & 3.01         & 2.83         & 2.77 \\
        & \(1{}^1\mathrm{B_{u}}\)  & 5.60 & 5.89 (6.5)  & 5.74 (8.1)   & 5.59 \\
        & \(2{}^1\mathrm{A_{g}}\)  & 6.55 &  4.21   & 5.73              & 5.58 \\
        \hline
        \hline
    \end{tabular}
    \begin{tablenotes}
    \item[a] Also shown are the energies from the semistochastic heat-bath CI (SHCI) method, extrapolated to the full CI limit.\cite{Chien:2018p2714} The $1s$ orbitals of carbon atoms were not included in the SHCI correlation treatment. The SHCI computations used the same basis sets and optimized geometries as those used for LR-OLCCD, LR-ODC-12, and EOM-CCSD.  
    \end{tablenotes}    
\end{threeparttable}
\end{table*}

Finally, we apply the LR-ODC-12 method to challenging excited states of ethylene
(\ce{C2H4}), butadiene (\ce{C4H6}), and hexatriene (\ce{C6H8}).
A reliable description of these electronic states requires an accurate treatment
of electron correlation.
\cite{Tavan:1986p6602,
Tavan:1987p4337,
Nakayama:1998p157,
Davidson:1996p6161,
Watts:1998p6979,
Muller:1999p7176,
Li:1999p177,
Starcke:2006p39,
Kurashige:2004p425,
Ghosh:2008p144117,
Sokolov:2017p244102,
Schreiber:2008p134110,
Zgid:2009p194107,
Angeli:2010p2436,
Daday:2012p4441,
Watson:2012p4013,
Zimmerman:2017p4712} 
All three molecules feature a dipole-allowed $1{}^1\mathrm{B_{u}}$ (or $1{}^1\mathrm{B_{1u}}$) state that is well described as a $\pi-\pi^*$ excitation, but requires a very accurate description of dynamic correlation between the $\sigma$ and $\pi$ electrons. In butadiene and hexatriene, the $1{}^1\mathrm{B_{u}}$ state is near-degenerate with a dipole-forbidden $2{}^1\mathrm{A_{g}}$ state that has a substantial double-excitation character, requiring the description of static correlation in the $\pi$ and $\pi^*$ orbitals.\cite{Kurashige:2004p425,Ghosh:2008p144117,Sokolov:2017p244102} For this reason, the relative energies and ordering of the $1{}^1\mathrm{B_{u}}$ and $2{}^1\mathrm{A_{g}}$ states are very sensitive to various levels of theory. For example, single-reference methods truncated to single and double excitations describe the $1{}^1\mathrm{B_{u}}$ state more accurately than the $2{}^1\mathrm{A_{g}}$ state, while multi-reference methods are more reliable for the $2{}^1\mathrm{A_{g}}$ state, missing important dynamic correlation for the $1{}^1\mathrm{B_{u}}$ state. Very recently, Chien et al.\cite{Chien:2018p2714} reported accurate vertical excitation energies for the low-lying states of ethylene, butadiene, and hexatriene computed using semistochastic heat-bath configuration interaction (SHCI) extrapolated to the full CI limit. In this section, we will use the SHCI results to benchmark the accuracy of the LR-ODC-12 method.

\cref{tab:alkenes} reports the ground-state total energies and vertical excitation energies of ethylene, butadiene, and hexatriene computed using the EOM-CCSD, LR-OLCCD, and LR-ODC-12 methods, along with the SHCI results from Ref.\@ \citenum{Chien:2018p2714}. We refer to the $\mathrm{B_{1u}}$ states of \ce{C2H4} as $\mathrm{B_{u}}$ for brevity. All methods employed the same optimized geometries and basis sets (see \cref{tab:alkenes} for details). We note that in the SHCI computations the $1s$ orbitals of carbon atoms were not included in the correlation treatment, while in other methods all electrons were correlated. To estimate the effect of the frozen-core approximation on the SHCI vertical excitation energies, we compared the excitation energies computed using the all-electron and frozen-core EOM-CCSD methods. The errors due to the frozen core did not exceed 0.01 eV.

All excitation energies decrease as the number of double bonds in a molecule increases. For butadiene and hexatriene, the ($1{}^1\mathrm{B_{u}}$; $2{}^1\mathrm{A_{g}}$) excitation energies computed using the SHCI method are (6.45; 6.58) and (5.59; 5.58) eV, respectively, indicating that the two states are nearly degenerate for the longer polyene. This feature is not reproduced by the EOM-CCSD method, which predicts the $1{}^1\mathrm{B_{u}}$ state energies in close agreement with SHCI, but significantly overestimates the energies for the doubly-excited $2{}^1\mathrm{A_{g}}$ state. As a result, the EOM-CCSD method overestimates the energy spacing between the $1{}^1\mathrm{B_{u}}$ and $2{}^1\mathrm{A_{g}}$ states by $\sim$ 0.6 eV and 1.0 eV for butadiene and hexatriene, respectively. 

The LR-ODC-12 method, by contrast, correctly describes the relative energies and
ordering of the $1{}^1\mathrm{B_{u}}$ and $2{}^1\mathrm{A_{g}}$ states. The
energy spacing between these states computed using LR-ODC-12 is 0.14 and $-$0.01
eV for butadiene and hexatriene, respectively, in an excellent agreement with
the SHCI results (0.13 and $-$0.01 eV). For the singlet excited states, the
LR-ODC-12 method consistently overestimates the excitation energies by $\sim$
0.1 -- 0.2 eV, relative to SHCI\@. For the $1{}^3\mathrm{B_{u}}$ state, the LR-ODC-12 errors are smaller in magnitude ($\sim$ 0.06 eV). Importantly, these results suggest that the LR-ODC-12 method provides a balanced description of the excited states with different electronic structure effects, as illustrated by its consistent performance for the $1{}^3\mathrm{B_{u}}$, $1{}^1\mathrm{B_{u}}$, and $2{}^1\mathrm{A_{g}}$ states in ethylene, butadiene, and hexatriene.

Comparing to LR-OLCCD shows that including the non-linear terms in LR-ODC-12 is crucial for the description of excited states with double-excitation character.
While for the $1{}^3\mathrm{B_{u}}$ and $1{}^1\mathrm{B_{u}}$ states the LR-OLCCD
errors exceed the LR-ODC-12 errors by $\sim$ 0.15 eV,
for the doubly-excited $2{}^1\mathrm{A_{g}}$ state the LR-OLCCD errors are
much bigger: 0.56 and $-$1.37 eV for butadiene and hexatriene, respectively.

\section{Conclusions}
\label{sec:conclusions}
We have presented a new approach for excited electronic states based on the linear-response formulation of density cumulant theory (DCT).
The resulting linear-response DCT model (LR-DCT) has the same computational
scaling as the original (single-state) DCT formulation but can accurately
predict energies and properties for many electronic states, simultaneously.
We have described the general formulation of LR-DCT, derived equations for the linear-response ODC-12 method (LR-ODC-12), and presented its implementation. In LR-ODC-12, excited-state energies are obtained by solving the generalized eigenvalue equation that involves a symmetric Hessian matrix.
This simplifies the computation of the excited-state properties (such as
transition dipoles) and ensures that the excitation energies have real values,
provided that the Hessian is positive semi-definite.
In addition, the LR-ODC-12 excitation energies are size-intensive, which we have
verified numerically for a system of noninteracting fragments. 

Our preliminary results demonstrate that LR-ODC-12 yields very accurate
excitation energies for a variety of excited states with different electronic
structure effects. For a set of small molecules (\ce{N2}, \ce{CO}, \ce{HCN},
\ce{HNC}, \ce{C2H2}, and \ce{H2CO}), LR-ODC-12 outperforms equation-of-motion
coupled cluster theory with single and double excitations (EOM-CCSD), with mean
absolute errors in excitation energies of less than 0.1 eV, relative to
reference data.
Importantly, both LR-ODC-12 and EOM-CCSD have the same computational scaling. In
a study of ethylene, butadiene, and hexatriene, we have compared the performance
of LR-ODC-12 and EOM-CCSD with the results from highly-accurate semistochastic
heat-bath configuration interaction (SHCI).
For
butadiene and hexatriene, LR-ODC-12 provides a balanced description of the
singly-excited $1{}^1\mathrm{B_{u}}$ and the doubly-excited
$2{}^1\mathrm{A_{g}}$ states, predicting that the two states become
nearly-degenerate in hexatriene, in excellent agreement with SHCI\@.
By contrast, EOM-CCSD drastically overestimates the energy of the
$2{}^1\mathrm{A_{g}}$ state, resulting in a $\sim$ 1 eV error in the energy gap
between these states of hexatriene.

Overall, our results demonstrate that linear-response density cumulant theory is
a promising theoretical approach for spectroscopic properties of molecules and
encourage its further development.
Several research directions are worth exploring. One of them is the efficient implementation of LR-ODC-12 and its applications to chemical systems with challenging electronic states. Two classes of systems that are particularly worth exploring are open-shell molecules and transition metal complexes. Another direction is to extend LR-DCT to simulations of other spectroscopic properties, such as photoelectron or X-ray absorption spectra.
In this regard, applying LR-DCT to the computation of optical rotation
properties is of particular interest as it is expected to avoid gauge invariance
problems due to the variational nature of the DCT orbitals.\cite{Lindh:2012p125}
We plan to explore these directions in the future.  

\appendix
\section{Derivatives of the One-Body Density Matrix in Density Cumulant Theory}
\label{sec:appendix}

Repeated differentiation of the one-body \(n\)-representability condition
(\cref{eq:one-body-n-rep}) gives the following formulas for the first and second
derivatives of the cumulant partial trace:
\begin{equation}
    \begin{array}{r@{\,}l}
        \dfrac{\partial \lambda^{pr}_{qr}}{\partial{y}}
        =
        \gamma^p_s
        \dfrac{\partial \gamma^s_q}{\partial{y}}
        +
        \dfrac{\partial \gamma^p_s}{\partial{y}}
        \gamma^s_q
        -
        \dfrac{\partial \gamma^p_q}{\partial{y}}
    \end{array}
\end{equation}
\begin{equation}
    \begin{array}{r@{\,}l}
        \dfrac{\partial^2 \lambda^{pr}_{qr}}{%
            \partial{x}
            \partial{y}
        }
        =
        &
        \gamma^p_s
        \dfrac{\partial \gamma^s_q}{%
            \partial{x}
            \partial{y}
        }
        +
        \dfrac{\partial \gamma^p_s}{%
            \partial{x} \partial{y}
        }
        \gamma^s_q
        -
        \dfrac{\partial \gamma^p_q}{%
            \partial{x}
            \partial{y}
        }
        \\[15pt]
        &
        +
        \dfrac{\partial \gamma^p_s}{\partial{x}}
        \dfrac{\partial \gamma^s_q}{\partial{y}}
        +
        \dfrac{\partial \gamma^s_q}{\partial{x}}
        \dfrac{\partial \gamma^p_s}{\partial{y}}
    \end{array}
\end{equation}
Transforming to the natural spin-orbital basis (NSO, denoted by prime indices) where the
one-body density matrix is diagonal, the first and second derivatives of the one-body density matrix can be determined from the cumulant derivatives as follows:
\begin{equation}
    \label{eq:one-body-n-rep-gradient}
    \begin{array}{r@{\,}l}
        \dfrac{\partial \gamma^{p'}_{q'}}{\partial{y}}
        =
        &
        \theta_{p'q'}
        \dfrac{\partial \lambda^{p'r}_{q'r}}{\partial{y}}
    \end{array}
\end{equation}
\begin{equation}
    \label{eq:one-body-n-rep-hessian}
    \begin{array}{r@{\,}l}
        \dfrac{\partial^2 \gamma^{p'}_{q'}}{%
            \partial{x}
            \partial{y}
        }
        =
        &
        \theta_{p'q'}
        \dfrac{\partial^2 \lambda^{p'r}_{q'r}}{%
            \partial{x}
            \partial{y}
        }
        -
        \delta_{r'}^{s'}
        \theta_{p'q'}
        \theta_{p's'}
        \theta_{q'r'}
        \dfrac{\partial \lambda^{p't}_{s't}}{%
            \partial{x}
        }
        \dfrac{\partial \lambda^{r'u}_{q'u}}{\partial{y}}
        \\[15pt]
        &
        -
        \delta_{r'}^{s'}
        \theta_{p'q'}
        \theta_{p's'}
        \theta_{q'r'}
        \dfrac{\partial \lambda^{r'u}_{q'u}}{\partial{x}}
        \dfrac{\partial \lambda^{p't}_{s't}}{\partial{y}}
    \end{array}
\end{equation}
Here, we have defined the following matrix:
\begin{equation}
    \theta_{p'q'}
    \equiv
    \left\{
        \begin{array}{cc}
            (
                \gamma_{p'}
                +
                \gamma_{q'}
                -
                1
            )^{-1}
            &
            \text{%
                if
                \(p',q'\in \mathrm{occ \ or \ vir}\)
            }
            \\
            0
            &
            \text{otherwise}
        \end{array}
    \right.
\end{equation}
where \(\gamma_{p'}\) denotes an eigenvalue of the one-body density matrix
(i.e., an occupation number). The natural spin-orbital $p'$ is considered occupied if \(\gamma_{p'} > 0.5\).

\cref{eq:one-body-n-rep-gradient,eq:one-body-n-rep-hessian} can be used to derive expressions for the two-body energy Hessian in \cref{eq:odc12-hessian-initial-form}. Simplifying the resulting equations allows us to determine the intermediates defined in \cref{eq:a22_odc12}. In the NSO basis, these intermediates are given by
\begin{equation}
    \label{eq:fancy-f-intermediate}
    \mathcal{F}_{p'}^{q'}
    \equiv
    \theta_{p'q'}
    f_{p'}^{q'}
\end{equation}
\begin{equation}
    \label{eq:fancy-g-intermediate}
    \mathcal{G}_{p'r'}^{q's'}
    \equiv
    \theta_{p'q'}
    \theta_{r's'}
    (
        \overline{g}_{p'r'}^{q's'}
        -
        \mathcal{F}_{p'}^{s'}
        \delta_{r'}^{q'}
        -
        \mathcal{F}_{r'}^{q'}
        \delta_{p'}^{s'}
    )
\end{equation}
These quantities are computed in the NSO basis and back-transformed to the
original spin-orbital basis using the eigenvectors of the one-particle density
matrix (see Ref.\@ \citenum{Sokolov:2013p024107} for more details).

\acknowledgement
A.V.C.\@ was supported by NSF grant CHE-1661604. A.Y.S.\@ was supported by start-up funds provided by the Ohio State University.

\suppinfo

Formulas for the energy Hessian (\(\mathbf{E}\)), the metric matrix
(\(\mathbf{M}\)), and the property gradient vector (\(\mathbf{v}'\)) for the
LR-ODC-12 and LR-OLCCD methods are included in the Supporting Information.

\providecommand{\latin}[1]{#1}
\makeatletter
\providecommand{\doi}
  {\begingroup\let\do\@makeother\dospecials
  \catcode`\{=1 \catcode`\}=2 \doi@aux}
\providecommand{\doi@aux}[1]{\endgroup\texttt{#1}}
\makeatother
\providecommand*\mcitethebibliography{\thebibliography}
\csname @ifundefined\endcsname{endmcitethebibliography}
  {\let\endmcitethebibliography\endthebibliography}{}


\begin{mcitethebibliography}{130}
\providecommand*\natexlab[1]{#1}
\providecommand*\mciteSetBstSublistMode[1]{}
\providecommand*\mciteSetBstMaxWidthForm[2]{}
\providecommand*\mciteBstWouldAddEndPuncttrue
  {\def\EndOfBibitem{\unskip.}}
\providecommand*\mciteBstWouldAddEndPunctfalse
  {\let\EndOfBibitem\relax}
\providecommand*\mciteSetBstMidEndSepPunct[3]{}
\providecommand*\mciteSetBstSublistLabelBeginEnd[3]{}
\providecommand*\EndOfBibitem{}
\mciteSetBstSublistMode{f}
\mciteSetBstMaxWidthForm{subitem}{(\alph{mcitesubitemcount})}
\mciteSetBstSublistLabelBeginEnd
  {\mcitemaxwidthsubitemform\space}
  {\relax}
  {\relax}

\bibitem[Knowles and Werner(1985)Knowles, and Werner]{Knowles:1985p259}
Knowles,~P.~J.; Werner,~H.-J. {An efficient second-order MC SCF method for long
  configuration expansions}. \emph{Chem. Phys. Lett.} \textbf{1985},
  \emph{115}, 259--267\relax
\mciteBstWouldAddEndPuncttrue
\mciteSetBstMidEndSepPunct{\mcitedefaultmidpunct}
{\mcitedefaultendpunct}{\mcitedefaultseppunct}\relax
\EndOfBibitem
\bibitem[Wolinski \latin{et~al.}(1987)Wolinski, Sellers, and
  Pulay]{Wolinski:1987p225}
Wolinski,~K.; Sellers,~H.~L.; Pulay,~P. {Consistent generalization of the
  M{\o}ller-Plesset partitioning to open-shell and multiconfigurational SCF
  reference states in many-body perturbation theory}. \emph{Chem. Phys. Lett.}
  \textbf{1987}, \emph{140}, 225--231\relax
\mciteBstWouldAddEndPuncttrue
\mciteSetBstMidEndSepPunct{\mcitedefaultmidpunct}
{\mcitedefaultendpunct}{\mcitedefaultseppunct}\relax
\EndOfBibitem
\bibitem[Hirao(1992)]{Hirao:1992p374}
Hirao,~K. {Multireference M{\o}ller{\textemdash}Plesset method}. \emph{Chem.
  Phys. Lett.} \textbf{1992}, \emph{190}, 374--380\relax
\mciteBstWouldAddEndPuncttrue
\mciteSetBstMidEndSepPunct{\mcitedefaultmidpunct}
{\mcitedefaultendpunct}{\mcitedefaultseppunct}\relax
\EndOfBibitem
\bibitem[Finley \latin{et~al.}(1998)Finley, Malmqvist, Roos, and
  Serrano-Andr{\'e}s]{Finley:1998p299}
Finley,~J.; Malmqvist,~P.~{\AA}.; Roos,~B.~O.; Serrano-Andr{\'e}s,~L. {The
  multi-state CASPT2 method}. \emph{Chem. Phys. Lett.} \textbf{1998},
  \emph{288}, 299--306\relax
\mciteBstWouldAddEndPuncttrue
\mciteSetBstMidEndSepPunct{\mcitedefaultmidpunct}
{\mcitedefaultendpunct}{\mcitedefaultseppunct}\relax
\EndOfBibitem
\bibitem[Andersson \latin{et~al.}(1990)Andersson, Malmqvist, Roos, Sadlej, and
  Wolinski]{Andersson:1990p5483}
Andersson,~K.; Malmqvist,~P.~{\AA}.; Roos,~B.~O.; Sadlej,~A.~J.; Wolinski,~K.
  {Second-order perturbation theory with a CASSCF reference function}. \emph{J.
  Phys. Chem.} \textbf{1990}, \emph{94}, 5483--5488\relax
\mciteBstWouldAddEndPuncttrue
\mciteSetBstMidEndSepPunct{\mcitedefaultmidpunct}
{\mcitedefaultendpunct}{\mcitedefaultseppunct}\relax
\EndOfBibitem
\bibitem[Andersson \latin{et~al.}(1992)Andersson, Malmqvist, and
  Roos]{Andersson:1992p1218}
Andersson,~K.; Malmqvist,~P.~{\AA}.; Roos,~B.~O. {Second-order perturbation
  theory with a complete active space self-consistent field reference
  function}. \emph{J. Chem. Phys.} \textbf{1992}, \emph{96}, 1218--1226\relax
\mciteBstWouldAddEndPuncttrue
\mciteSetBstMidEndSepPunct{\mcitedefaultmidpunct}
{\mcitedefaultendpunct}{\mcitedefaultseppunct}\relax
\EndOfBibitem
\bibitem[Angeli \latin{et~al.}(2001)Angeli, Cimiraglia, Evangelisti, Leininger,
  and Malrieu]{Angeli:2001p10252}
Angeli,~C.; Cimiraglia,~R.; Evangelisti,~S.; Leininger,~T.; Malrieu,~J.-P.~P.
  {Introduction of n-electron valence states for multireference perturbation
  theory}. \emph{J. Chem. Phys.} \textbf{2001}, \emph{114}, 10252--10264\relax
\mciteBstWouldAddEndPuncttrue
\mciteSetBstMidEndSepPunct{\mcitedefaultmidpunct}
{\mcitedefaultendpunct}{\mcitedefaultseppunct}\relax
\EndOfBibitem
\bibitem[Angeli \latin{et~al.}(2001)Angeli, Cimiraglia, and
  Malrieu]{Angeli:2001p297}
Angeli,~C.; Cimiraglia,~R.; Malrieu,~J.-P.~P. {N-electron valence state
  perturbation theory: a fast implementation of the strongly contracted
  variant}. \emph{Chem. Phys. Lett.} \textbf{2001}, \emph{350}, 297--305\relax
\mciteBstWouldAddEndPuncttrue
\mciteSetBstMidEndSepPunct{\mcitedefaultmidpunct}
{\mcitedefaultendpunct}{\mcitedefaultseppunct}\relax
\EndOfBibitem
\bibitem[Mukherjee \latin{et~al.}(1977)Mukherjee, Moitra, and
  Mukhopadhyay]{Mukherjee:1977p955}
Mukherjee,~D.; Moitra,~R.~K.; Mukhopadhyay,~A. {Applications of a
  non-perturbative many-body formalism to general open-shell atomic and
  molecular problems: calculation of the ground and the lowest $\pi$-$\pi$*
  singlet and triplet energies and the first ionization potential of
  trans-butadiene}. \emph{Mol. Phys.} \textbf{1977}, \emph{33}, 955--969\relax
\mciteBstWouldAddEndPuncttrue
\mciteSetBstMidEndSepPunct{\mcitedefaultmidpunct}
{\mcitedefaultendpunct}{\mcitedefaultseppunct}\relax
\EndOfBibitem
\bibitem[Jeziorski and Monkhorst(1981)Jeziorski, and
  Monkhorst]{Jeziorski:1981p1668}
Jeziorski,~B.; Monkhorst,~H.~J. {Coupled-cluster method for multideterminantal
  reference states}. \emph{Phys. Rev. A} \textbf{1981}, \emph{24},
  1668--1681\relax
\mciteBstWouldAddEndPuncttrue
\mciteSetBstMidEndSepPunct{\mcitedefaultmidpunct}
{\mcitedefaultendpunct}{\mcitedefaultseppunct}\relax
\EndOfBibitem
\bibitem[Werner and Knowles(1988)Werner, and Knowles]{Werner:1988p5803}
Werner,~H.-J.; Knowles,~P.~J. {An efficient internally contracted
  multiconfiguration{\textendash}reference configuration interaction method}.
  \emph{J. Chem. Phys.} \textbf{1988}, \emph{89}, 5803--5814\relax
\mciteBstWouldAddEndPuncttrue
\mciteSetBstMidEndSepPunct{\mcitedefaultmidpunct}
{\mcitedefaultendpunct}{\mcitedefaultseppunct}\relax
\EndOfBibitem
\bibitem[Mahapatra \latin{et~al.}(1998)Mahapatra, Datta, and
  Mukherjee]{Mahapatra:1998p157}
Mahapatra,~U.~S.; Datta,~B.; Mukherjee,~D. {A state-specific multi-reference
  coupled cluster formalism with molecular applications}. \emph{Mol. Phys.}
  \textbf{1998}, \emph{94}, 157--171\relax
\mciteBstWouldAddEndPuncttrue
\mciteSetBstMidEndSepPunct{\mcitedefaultmidpunct}
{\mcitedefaultendpunct}{\mcitedefaultseppunct}\relax
\EndOfBibitem
\bibitem[Pittner(2003)]{Pittner:2003p10876}
Pittner,~J. {Continuous transition between Brillouin{\textendash}Wigner and
  Rayleigh{\textendash}Schr{\"o}dinger perturbation theory, generalized Bloch
  equation, and Hilbert space multireference coupled cluster}. \emph{J. Chem.
  Phys.} \textbf{2003}, \emph{118}, 10876\relax
\mciteBstWouldAddEndPuncttrue
\mciteSetBstMidEndSepPunct{\mcitedefaultmidpunct}
{\mcitedefaultendpunct}{\mcitedefaultseppunct}\relax
\EndOfBibitem
\bibitem[Evangelista \latin{et~al.}(2007)Evangelista, Allen, and
  Schaefer]{Evangelista:2007p024102}
Evangelista,~F.~A.; Allen,~W.~D.; Schaefer,~H.~F. {Coupling term derivation and
  general implementation of state-specific multireference coupled cluster
  theories}. \emph{J. Chem. Phys.} \textbf{2007}, \emph{127}, 024102\relax
\mciteBstWouldAddEndPuncttrue
\mciteSetBstMidEndSepPunct{\mcitedefaultmidpunct}
{\mcitedefaultendpunct}{\mcitedefaultseppunct}\relax
\EndOfBibitem
\bibitem[Datta \latin{et~al.}(2011)Datta, Kong, and Nooijen]{Datta:2011p214116}
Datta,~D.; Kong,~L.; Nooijen,~M. {A state-specific partially internally
  contracted multireference coupled cluster approach}. \emph{J. Chem. Phys.}
  \textbf{2011}, \emph{134}, 214116--214116--19\relax
\mciteBstWouldAddEndPuncttrue
\mciteSetBstMidEndSepPunct{\mcitedefaultmidpunct}
{\mcitedefaultendpunct}{\mcitedefaultseppunct}\relax
\EndOfBibitem
\bibitem[Evangelista and Gauss(2011)Evangelista, and
  Gauss]{Evangelista:2011p114102}
Evangelista,~F.~A.; Gauss,~J. {An orbital-invariant internally contracted
  multireference coupled cluster approach}. \emph{J. Chem. Phys.}
  \textbf{2011}, \emph{134}, 114102\relax
\mciteBstWouldAddEndPuncttrue
\mciteSetBstMidEndSepPunct{\mcitedefaultmidpunct}
{\mcitedefaultendpunct}{\mcitedefaultseppunct}\relax
\EndOfBibitem
\bibitem[K{\"o}hn \latin{et~al.}(2013)K{\"o}hn, Hanauer, M{\"u}ck, Jagau, and
  Gauss]{Kohn:2012p176}
K{\"o}hn,~A.; Hanauer,~M.; M{\"u}ck,~L.~A.; Jagau,~T.-C.; Gauss,~J.
  {State-specific multireference coupled-cluster theory}. \emph{WIREs Comput.
  Mol. Sci.} \textbf{2013}, \emph{3}, 176--197\relax
\mciteBstWouldAddEndPuncttrue
\mciteSetBstMidEndSepPunct{\mcitedefaultmidpunct}
{\mcitedefaultendpunct}{\mcitedefaultseppunct}\relax
\EndOfBibitem
\bibitem[Nooijen \latin{et~al.}(2014)Nooijen, Demel, Datta, Kong, Shamasundar,
  Lotrich, Huntington, and Neese]{Nooijen:2014p081102}
Nooijen,~M.; Demel,~O.; Datta,~D.; Kong,~L.; Shamasundar,~K.~R.; Lotrich,~V.;
  Huntington,~L.~M.; Neese,~F. {Communication: Multireference equation of
  motion coupled cluster: A transform and diagonalize approach to electronic
  structure}. \emph{J. Chem. Phys.} \textbf{2014}, \emph{140}, 081102\relax
\mciteBstWouldAddEndPuncttrue
\mciteSetBstMidEndSepPunct{\mcitedefaultmidpunct}
{\mcitedefaultendpunct}{\mcitedefaultseppunct}\relax
\EndOfBibitem
\bibitem[Foresman \latin{et~al.}(1992)Foresman, Head-Gordon, Pople, and
  Frisch]{Foresman:1992p135}
Foresman,~J.~B.; Head-Gordon,~M.; Pople,~J.~A.; Frisch,~M.~J. {Toward a
  systematic molecular orbital theory for excited states}. \emph{J. Phys.
  Chem.} \textbf{1992}, \emph{96}, 135--149\relax
\mciteBstWouldAddEndPuncttrue
\mciteSetBstMidEndSepPunct{\mcitedefaultmidpunct}
{\mcitedefaultendpunct}{\mcitedefaultseppunct}\relax
\EndOfBibitem
\bibitem[Sherrill and Schaefer(1999)Sherrill, and Schaefer]{Sherrill:1999p143}
Sherrill,~C.~D.; Schaefer,~H.~F. {The Configuration Interaction Method:
  Advances in Highly Correlated Approaches}. \emph{Adv. Quant. Chem.}
  \textbf{1999}, \emph{34}, 143--269\relax
\mciteBstWouldAddEndPuncttrue
\mciteSetBstMidEndSepPunct{\mcitedefaultmidpunct}
{\mcitedefaultendpunct}{\mcitedefaultseppunct}\relax
\EndOfBibitem
\bibitem[Geertsen \latin{et~al.}(1989)Geertsen, Rittby, and
  Bartlett]{Geertsen:1989p57}
Geertsen,~J.; Rittby,~M.; Bartlett,~R.~J. {The equation-of-motion
  coupled-cluster method: Excitation energies of Be and CO}. \emph{Chem. Phys.
  Lett.} \textbf{1989}, \emph{164}, 57--62\relax
\mciteBstWouldAddEndPuncttrue
\mciteSetBstMidEndSepPunct{\mcitedefaultmidpunct}
{\mcitedefaultendpunct}{\mcitedefaultseppunct}\relax
\EndOfBibitem
\bibitem[Comeau and Bartlett(1993)Comeau, and Bartlett]{Comeau:1993p414}
Comeau,~D.~C.; Bartlett,~R.~J. {The equation-of-motion coupled-cluster method.
  Applications to open- and closed-shell reference states}. \emph{Chem. Phys.
  Lett.} \textbf{1993}, \emph{207}, 414--423\relax
\mciteBstWouldAddEndPuncttrue
\mciteSetBstMidEndSepPunct{\mcitedefaultmidpunct}
{\mcitedefaultendpunct}{\mcitedefaultseppunct}\relax
\EndOfBibitem
\bibitem[Stanton and Bartlett(1993)Stanton, and Bartlett]{Stanton:1993p7029}
Stanton,~J.~F.; Bartlett,~R.~J. {The equation of motion coupled-cluster method.
  A systematic biorthogonal approach to molecular excitation energies,
  transition probabilities, and excited state properties}. \emph{J. Chem.
  Phys.} \textbf{1993}, \emph{98}, 7029\relax
\mciteBstWouldAddEndPuncttrue
\mciteSetBstMidEndSepPunct{\mcitedefaultmidpunct}
{\mcitedefaultendpunct}{\mcitedefaultseppunct}\relax
\EndOfBibitem
\bibitem[Krylov(2008)]{Krylov:2008p433}
Krylov,~A.~I. {Equation-of-Motion Coupled-Cluster Methods for Open-Shell and
  Electronically Excited Species: The Hitchhiker's Guide to Fock Space}.
  \emph{Annu. Rev. Phys. Chem.} \textbf{2008}, \emph{59}, 433--462\relax
\mciteBstWouldAddEndPuncttrue
\mciteSetBstMidEndSepPunct{\mcitedefaultmidpunct}
{\mcitedefaultendpunct}{\mcitedefaultseppunct}\relax
\EndOfBibitem
\bibitem[Crawford and Schaefer(2000)Crawford, and Schaefer]{Crawford:2000p33}
Crawford,~T.~D.; Schaefer,~H.~F. {An Introduction to Coupled Cluster Theory for
  Computational Chemists}. \emph{Rev. Comp. Chem.} \textbf{2000}, \emph{14},
  33--136\relax
\mciteBstWouldAddEndPuncttrue
\mciteSetBstMidEndSepPunct{\mcitedefaultmidpunct}
{\mcitedefaultendpunct}{\mcitedefaultseppunct}\relax
\EndOfBibitem
\bibitem[Shavitt and Bartlett(2009)Shavitt, and Bartlett]{Shavitt:2009}
Shavitt,~I.; Bartlett,~R.~J. \emph{Many-Body Methods in Chemistry and Physics};
  Cambridge University Press: Cambridge, UK, 2009\relax
\mciteBstWouldAddEndPuncttrue
\mciteSetBstMidEndSepPunct{\mcitedefaultmidpunct}
{\mcitedefaultendpunct}{\mcitedefaultseppunct}\relax
\EndOfBibitem
\bibitem[Sekino and Bartlett(1984)Sekino, and Bartlett]{Sekino:1984p255}
Sekino,~H.; Bartlett,~R.~J. {A linear response, coupled-cluster theory for
  excitation energy}. \emph{Int. J. Quantum Chem.} \textbf{1984}, \emph{26},
  255--265\relax
\mciteBstWouldAddEndPuncttrue
\mciteSetBstMidEndSepPunct{\mcitedefaultmidpunct}
{\mcitedefaultendpunct}{\mcitedefaultseppunct}\relax
\EndOfBibitem
\bibitem[Koch \latin{et~al.}(1990)Koch, Jensen, J{\o}rgensen, and
  Helgaker]{Koch:1990p3345}
Koch,~H.; Jensen,~H. J.~A.; J{\o}rgensen,~P.; Helgaker,~T. {Excitation energies
  from the coupled cluster singles and doubles linear response function
  (CCSDLR). Applications to Be, CH+, CO, and H2O}. \emph{J. Chem. Phys.}
  \textbf{1990}, \emph{93}, 3345--3350\relax
\mciteBstWouldAddEndPuncttrue
\mciteSetBstMidEndSepPunct{\mcitedefaultmidpunct}
{\mcitedefaultendpunct}{\mcitedefaultseppunct}\relax
\EndOfBibitem
\bibitem[Koch and J{\o}rgensen(1990)Koch, and J{\o}rgensen]{Koch:1990p3333}
Koch,~H.; J{\o}rgensen,~P. {Coupled cluster response functions}. \emph{J. Chem.
  Phys.} \textbf{1990}, \emph{93}, 3333--3344\relax
\mciteBstWouldAddEndPuncttrue
\mciteSetBstMidEndSepPunct{\mcitedefaultmidpunct}
{\mcitedefaultendpunct}{\mcitedefaultseppunct}\relax
\EndOfBibitem
\bibitem[Nooijen and Bartlett(1997)Nooijen, and Bartlett]{Nooijen:1997p6441}
Nooijen,~M.; Bartlett,~R.~J. {A new method for excited states: Similarity
  transformed equation-of-motion coupled-cluster theory}. \emph{J. Chem. Phys.}
  \textbf{1997}, \emph{106}, 6441--6448\relax
\mciteBstWouldAddEndPuncttrue
\mciteSetBstMidEndSepPunct{\mcitedefaultmidpunct}
{\mcitedefaultendpunct}{\mcitedefaultseppunct}\relax
\EndOfBibitem
\bibitem[Nooijen and Bartlett(1997)Nooijen, and Bartlett]{Nooijen:1997p6812}
Nooijen,~M.; Bartlett,~R.~J. {Similarity transformed equation-of-motion
  coupled-cluster theory: Details, examples, and comparisons}. \emph{J. Chem.
  Phys.} \textbf{1997}, \emph{107}, 6812--6830\relax
\mciteBstWouldAddEndPuncttrue
\mciteSetBstMidEndSepPunct{\mcitedefaultmidpunct}
{\mcitedefaultendpunct}{\mcitedefaultseppunct}\relax
\EndOfBibitem
\bibitem[Nakatsuji and Hirao(1978)Nakatsuji, and Hirao]{Nakatsuji:1978p2053}
Nakatsuji,~H.; Hirao,~K. {Cluster expansion of the wavefunction.
  Symmetry-adapted-cluster expansion, its variational determination, and
  extension of open-shell orbital theory}. \emph{J. Chem. Phys.} \textbf{1978},
  \emph{68}, 2053--2065\relax
\mciteBstWouldAddEndPuncttrue
\mciteSetBstMidEndSepPunct{\mcitedefaultmidpunct}
{\mcitedefaultendpunct}{\mcitedefaultseppunct}\relax
\EndOfBibitem
\bibitem[Nakatsuji(1979)]{Nakatsuji:1979p329}
Nakatsuji,~H. {Cluster expansion of the wavefunction. Electron correlations in
  ground and excited states by SAC (symmetry-adapted-cluster) and SAC CI
  theories}. \emph{Chem. Phys. Lett.} \textbf{1979}, \emph{67}, 329--333\relax
\mciteBstWouldAddEndPuncttrue
\mciteSetBstMidEndSepPunct{\mcitedefaultmidpunct}
{\mcitedefaultendpunct}{\mcitedefaultseppunct}\relax
\EndOfBibitem
\bibitem[Stanton(1993)]{Stanton:1993p8840}
Stanton,~J.~F. {Many-body methods for excited state potential energy surfaces.
  I. General theory of energy gradients for the equation-of-motion
  coupled-cluster method}. \emph{J. Chem. Phys.} \textbf{1993}, \emph{99},
  8840--8847\relax
\mciteBstWouldAddEndPuncttrue
\mciteSetBstMidEndSepPunct{\mcitedefaultmidpunct}
{\mcitedefaultendpunct}{\mcitedefaultseppunct}\relax
\EndOfBibitem
\bibitem[Stanton and Gauss(1994)Stanton, and Gauss]{Stanton:1994p4695}
Stanton,~J.~F.; Gauss,~J. {Analytic energy gradients for the equation-of-motion
  coupled-cluster method: Implementation and application to the HCN/HNC
  system}. \emph{J. Chem. Phys.} \textbf{1994}, \emph{100}, 4695--4698\relax
\mciteBstWouldAddEndPuncttrue
\mciteSetBstMidEndSepPunct{\mcitedefaultmidpunct}
{\mcitedefaultendpunct}{\mcitedefaultseppunct}\relax
\EndOfBibitem
\bibitem[Stanton and Gauss(1994)Stanton, and Gauss]{Stanton:1994p8938}
Stanton,~J.~F.; Gauss,~J. {Analytic energy derivatives for ionized states
  described by the equation-of-motion coupled cluster method}. \emph{J. Chem.
  Phys.} \textbf{1994}, \emph{101}, 8938--8944\relax
\mciteBstWouldAddEndPuncttrue
\mciteSetBstMidEndSepPunct{\mcitedefaultmidpunct}
{\mcitedefaultendpunct}{\mcitedefaultseppunct}\relax
\EndOfBibitem
\bibitem[Levchenko \latin{et~al.}(2005)Levchenko, Wang, and
  Krylov]{Levchenko:2005p224106}
Levchenko,~S.~V.; Wang,~T.; Krylov,~A.~I. {Analytic gradients for the
  spin-conserving and spin-flipping equation-of-motion coupled-cluster models
  with single and double substitutions}. \emph{J. Chem. Phys.} \textbf{2005},
  \emph{122}, 224106\relax
\mciteBstWouldAddEndPuncttrue
\mciteSetBstMidEndSepPunct{\mcitedefaultmidpunct}
{\mcitedefaultendpunct}{\mcitedefaultseppunct}\relax
\EndOfBibitem
\bibitem[H{\"a}ttig(2005)]{Hattig:2005p37}
H{\"a}ttig,~C. Structure Optimizations for Excited States with Correlated
  Second-Order Methods: CC2 and ADC(2). \emph{Adv. Quant. Chem.} \textbf{2005},
  \emph{50}, 37 -- 60\relax
\mciteBstWouldAddEndPuncttrue
\mciteSetBstMidEndSepPunct{\mcitedefaultmidpunct}
{\mcitedefaultendpunct}{\mcitedefaultseppunct}\relax
\EndOfBibitem
\bibitem[K{\"o}hn and Tajti(2007)K{\"o}hn, and Tajti]{Kohn:2007p044105}
K{\"o}hn,~A.; Tajti,~A. {Can coupled-cluster theory treat conical
  intersections?} \emph{J. Chem. Phys.} \textbf{2007}, \emph{127}, 044105\relax
\mciteBstWouldAddEndPuncttrue
\mciteSetBstMidEndSepPunct{\mcitedefaultmidpunct}
{\mcitedefaultendpunct}{\mcitedefaultseppunct}\relax
\EndOfBibitem
\bibitem[Kj{\o}nstad \latin{et~al.}(2017)Kj{\o}nstad, Myhre, Mart{\'\i}nez, and
  Koch]{Kjonstad:2017p164105}
Kj{\o}nstad,~E.~F.; Myhre,~R.~H.; Mart{\'\i}nez,~T.~J.; Koch,~H. {Crossing
  conditions in coupled cluster theory}. \emph{J. Chem. Phys.} \textbf{2017},
  \emph{147}, 164105\relax
\mciteBstWouldAddEndPuncttrue
\mciteSetBstMidEndSepPunct{\mcitedefaultmidpunct}
{\mcitedefaultendpunct}{\mcitedefaultseppunct}\relax
\EndOfBibitem
\bibitem[Schirmer(1982)]{Schirmer:1982p2395}
Schirmer,~J. {Beyond the random-phase approximation: A new approximation scheme
  for the polarization propagator}. \emph{Phys. Rev. A} \textbf{1982},
  \emph{26}, 2395--2416\relax
\mciteBstWouldAddEndPuncttrue
\mciteSetBstMidEndSepPunct{\mcitedefaultmidpunct}
{\mcitedefaultendpunct}{\mcitedefaultseppunct}\relax
\EndOfBibitem
\bibitem[Schirmer(1991)]{Schirmer:1991p4647}
Schirmer,~J. {Closed-form intermediate representations of many-body propagators
  and resolvent matrices}. \emph{Phys. Rev. A} \textbf{1991}, \emph{43},
  4647\relax
\mciteBstWouldAddEndPuncttrue
\mciteSetBstMidEndSepPunct{\mcitedefaultmidpunct}
{\mcitedefaultendpunct}{\mcitedefaultseppunct}\relax
\EndOfBibitem
\bibitem[Dreuw and Wormit(2014)Dreuw, and Wormit]{Dreuw:2014p82}
Dreuw,~A.; Wormit,~M. {The algebraic diagrammatic construction scheme for the
  polarization propagator for the calculation of excited states}. \emph{WIREs
  Comput. Mol. Sci.} \textbf{2014}, \emph{5}, 82--95\relax
\mciteBstWouldAddEndPuncttrue
\mciteSetBstMidEndSepPunct{\mcitedefaultmidpunct}
{\mcitedefaultendpunct}{\mcitedefaultseppunct}\relax
\EndOfBibitem
\bibitem[Taube and Bartlett(2006)Taube, and Bartlett]{Taube:2006p3393}
Taube,~A.~G.; Bartlett,~R.~J. {New perspectives on unitary coupled-cluster
  theory}. \emph{Int. J. Quantum Chem.} \textbf{2006}, \emph{106},
  3393--3401\relax
\mciteBstWouldAddEndPuncttrue
\mciteSetBstMidEndSepPunct{\mcitedefaultmidpunct}
{\mcitedefaultendpunct}{\mcitedefaultseppunct}\relax
\EndOfBibitem
\bibitem[Kats \latin{et~al.}(2011)Kats, Usvyat, and
  Sch{\"u}tz]{Kats:2011p062503}
Kats,~D.; Usvyat,~D.; Sch{\"u}tz,~M. {Second-order variational coupled-cluster
  linear-response method: A Hermitian time-dependent theory}. \emph{Phys. Rev.
  A} \textbf{2011}, \emph{83}, 062503\relax
\mciteBstWouldAddEndPuncttrue
\mciteSetBstMidEndSepPunct{\mcitedefaultmidpunct}
{\mcitedefaultendpunct}{\mcitedefaultseppunct}\relax
\EndOfBibitem
\bibitem[W{\"a}lz \latin{et~al.}(2012)W{\"a}lz, Kats, Usvyat, Korona, and
  Sch{\"u}tz]{Walz:2012p052519}
W{\"a}lz,~G.; Kats,~D.; Usvyat,~D.; Korona,~T.; Sch{\"u}tz,~M. {Application of
  Hermitian time-dependent coupled-cluster response Ans{\"a}tze of second order
  to excitation energies and frequency-dependent dipole polarizabilities}.
  \emph{Phys. Rev. A} \textbf{2012}, \emph{86}, 052519\relax
\mciteBstWouldAddEndPuncttrue
\mciteSetBstMidEndSepPunct{\mcitedefaultmidpunct}
{\mcitedefaultendpunct}{\mcitedefaultseppunct}\relax
\EndOfBibitem
\bibitem[Kj{\o}nstad and Koch(2017)Kj{\o}nstad, and Koch]{Kjonstad:2017p4801}
Kj{\o}nstad,~E.~F.; Koch,~H. {Resolving the Notorious Case of Conical
  Intersections for Coupled Cluster Dynamics}. \emph{J. Phys. Chem. Lett.}
  \textbf{2017}, \emph{8}, 4801--4807\relax
\mciteBstWouldAddEndPuncttrue
\mciteSetBstMidEndSepPunct{\mcitedefaultmidpunct}
{\mcitedefaultendpunct}{\mcitedefaultseppunct}\relax
\EndOfBibitem
\bibitem[Moszynski \latin{et~al.}(2005)Moszynski, {\.{Z}}uchowski, and
  Jeziorski]{Moszynski:2005p1109}
Moszynski,~R.; {\.{Z}}uchowski,~P.~S.; Jeziorski,~B. {Time-Independent
  Coupled-Cluster Theory of the Polarization Propagator}. \emph{Collect. Czech.
  Chem. Commun.} \textbf{2005}, \emph{70}, 1109--1132\relax
\mciteBstWouldAddEndPuncttrue
\mciteSetBstMidEndSepPunct{\mcitedefaultmidpunct}
{\mcitedefaultendpunct}{\mcitedefaultseppunct}\relax
\EndOfBibitem
\bibitem[Korona(2010)]{Korona:2010p14977}
Korona,~T. {XCC2{\textemdash}a new coupled cluster model for the second-order
  polarization propagator}. \emph{Phys. Chem. Chem. Phys.} \textbf{2010},
  \emph{12}, 14977--14984\relax
\mciteBstWouldAddEndPuncttrue
\mciteSetBstMidEndSepPunct{\mcitedefaultmidpunct}
{\mcitedefaultendpunct}{\mcitedefaultseppunct}\relax
\EndOfBibitem
\bibitem[Kutzelnigg(2006)]{Kutzelnigg:2006p171101}
Kutzelnigg,~W. {Density-cumulant functional theory}. \emph{J. Chem. Phys.}
  \textbf{2006}, \emph{125}, 171101\relax
\mciteBstWouldAddEndPuncttrue
\mciteSetBstMidEndSepPunct{\mcitedefaultmidpunct}
{\mcitedefaultendpunct}{\mcitedefaultseppunct}\relax
\EndOfBibitem
\bibitem[Simmonett \latin{et~al.}(2010)Simmonett, Wilke, Schaefer, and
  Kutzelnigg]{Simmonett:2010p174122}
Simmonett,~A.~C.; Wilke,~J.~J.; Schaefer,~H.~F.; Kutzelnigg,~W. {Density
  cumulant functional theory: First implementation and benchmark results for
  the DCFT-06 model}. \emph{J. Chem. Phys.} \textbf{2010}, \emph{133},
  174122\relax
\mciteBstWouldAddEndPuncttrue
\mciteSetBstMidEndSepPunct{\mcitedefaultmidpunct}
{\mcitedefaultendpunct}{\mcitedefaultseppunct}\relax
\EndOfBibitem
\bibitem[Sokolov \latin{et~al.}(2012)Sokolov, Wilke, Simmonett, and
  Schaefer]{Sokolov:2012p054105}
Sokolov,~A.~Y.; Wilke,~J.~J.; Simmonett,~A.~C.; Schaefer,~H.~F. {Analytic
  gradients for density cumulant functional theory: The DCFT-06 model}.
  \emph{J. Chem. Phys.} \textbf{2012}, \emph{137}, 054105--054105--7\relax
\mciteBstWouldAddEndPuncttrue
\mciteSetBstMidEndSepPunct{\mcitedefaultmidpunct}
{\mcitedefaultendpunct}{\mcitedefaultseppunct}\relax
\EndOfBibitem
\bibitem[Sokolov \latin{et~al.}(2013)Sokolov, Simmonett, and
  Schaefer]{Sokolov:2013p024107}
Sokolov,~A.~Y.; Simmonett,~A.~C.; Schaefer,~H.~F. {Density cumulant functional
  theory: The DC-12 method, an improved description of the one-particle density
  matrix}. \emph{J. Chem. Phys.} \textbf{2013}, \emph{138},
  024107--024107--9\relax
\mciteBstWouldAddEndPuncttrue
\mciteSetBstMidEndSepPunct{\mcitedefaultmidpunct}
{\mcitedefaultendpunct}{\mcitedefaultseppunct}\relax
\EndOfBibitem
\bibitem[Sokolov and Schaefer(2013)Sokolov, and Schaefer]{Sokolov:2013p204110}
Sokolov,~A.~Y.; Schaefer,~H.~F. {Orbital-optimized density cumulant functional
  theory}. \emph{J. Chem. Phys.} \textbf{2013}, \emph{139},
  204110--204110\relax
\mciteBstWouldAddEndPuncttrue
\mciteSetBstMidEndSepPunct{\mcitedefaultmidpunct}
{\mcitedefaultendpunct}{\mcitedefaultseppunct}\relax
\EndOfBibitem
\bibitem[Sokolov \latin{et~al.}(2014)Sokolov, Schaefer, and
  Kutzelnigg]{Sokolov:2014p074111}
Sokolov,~A.~Y.; Schaefer,~H.~F.; Kutzelnigg,~W. {Density cumulant functional
  theory from a unitary transformation: N-representability, three-particle
  correlation effects, and application to O4+}. \emph{J. Chem. Phys.}
  \textbf{2014}, \emph{141}, 074111\relax
\mciteBstWouldAddEndPuncttrue
\mciteSetBstMidEndSepPunct{\mcitedefaultmidpunct}
{\mcitedefaultendpunct}{\mcitedefaultseppunct}\relax
\EndOfBibitem
\bibitem[Wang \latin{et~al.}(2016)Wang, Sokolov, Turney, and
  Schaefer]{Wang:2016p4833}
Wang,~X.; Sokolov,~A.~Y.; Turney,~J.~M.; Schaefer,~H.~F. {Spin-Adapted
  Formulation and Implementation of Density Cumulant Functional Theory with
  Density-Fitting Approximation: Application to Transition Metal Compounds}.
  \emph{J. Chem. Theory Comput.} \textbf{2016}, \emph{12}, 4833--4842\relax
\mciteBstWouldAddEndPuncttrue
\mciteSetBstMidEndSepPunct{\mcitedefaultmidpunct}
{\mcitedefaultendpunct}{\mcitedefaultseppunct}\relax
\EndOfBibitem
\bibitem[DCF()]{DCFTfootnote}
We note that early on DCT was referred to as ``density cumulant functional
  theory'' (DCFT).\relax
\mciteBstWouldAddEndPunctfalse
\mciteSetBstMidEndSepPunct{\mcitedefaultmidpunct}
{}{\mcitedefaultseppunct}\relax
\EndOfBibitem
\bibitem[Fulde(1991)]{Fulde:1991}
Fulde,~P. \emph{{Electron Correlations in Molecules and Solids}}; Springer:
  Berlin, 1991\relax
\mciteBstWouldAddEndPuncttrue
\mciteSetBstMidEndSepPunct{\mcitedefaultmidpunct}
{\mcitedefaultendpunct}{\mcitedefaultseppunct}\relax
\EndOfBibitem
\bibitem[Ziesche(1992)]{Ziesche:1992p597}
Ziesche,~P. {Definition of exchange based on cumulant expansion: Correlation
  induced narrowing of the exchange hole}. \emph{Solid State Commun}
  \textbf{1992}, \emph{82}, 597--602\relax
\mciteBstWouldAddEndPuncttrue
\mciteSetBstMidEndSepPunct{\mcitedefaultmidpunct}
{\mcitedefaultendpunct}{\mcitedefaultseppunct}\relax
\EndOfBibitem
\bibitem[Kutzelnigg and Mukherjee(1997)Kutzelnigg, and
  Mukherjee]{Kutzelnigg:1997p432}
Kutzelnigg,~W.; Mukherjee,~D. {Normal order and extended Wick theorem for a
  multiconfiguration reference wave function}. \emph{J. Chem. Phys.}
  \textbf{1997}, \emph{107}, 432\relax
\mciteBstWouldAddEndPuncttrue
\mciteSetBstMidEndSepPunct{\mcitedefaultmidpunct}
{\mcitedefaultendpunct}{\mcitedefaultseppunct}\relax
\EndOfBibitem
\bibitem[Mazziotti(1998)]{Mazziotti:1998p419}
Mazziotti,~D.~A. {Approximate solution for electron correlation through the use
  of Schwinger probes}. \emph{Chem. Phys. Lett.} \textbf{1998}, \emph{289},
  419--427\relax
\mciteBstWouldAddEndPuncttrue
\mciteSetBstMidEndSepPunct{\mcitedefaultmidpunct}
{\mcitedefaultendpunct}{\mcitedefaultseppunct}\relax
\EndOfBibitem
\bibitem[Mazziotti(1998)]{Mazziotti:1998p4219}
Mazziotti,~D.~A. {Contracted Schr{\"o}dinger equation: Determining quantum
  energies and two-particle density matrices without wave functions}.
  \emph{Phys. Rev. A} \textbf{1998}, \emph{57}, 4219--4234\relax
\mciteBstWouldAddEndPuncttrue
\mciteSetBstMidEndSepPunct{\mcitedefaultmidpunct}
{\mcitedefaultendpunct}{\mcitedefaultseppunct}\relax
\EndOfBibitem
\bibitem[Kutzelnigg and Mukherjee(1999)Kutzelnigg, and
  Mukherjee]{Kutzelnigg:1999p2800}
Kutzelnigg,~W.; Mukherjee,~D. {Cumulant expansion of the reduced density
  matrices}. \emph{J. Chem. Phys.} \textbf{1999}, \emph{110}, 2800--2809\relax
\mciteBstWouldAddEndPuncttrue
\mciteSetBstMidEndSepPunct{\mcitedefaultmidpunct}
{\mcitedefaultendpunct}{\mcitedefaultseppunct}\relax
\EndOfBibitem
\bibitem[Ziesche(2000)]{Ziesche:2000p33}
Ziesche,~P. In \emph{Many-Electron Densities and Reduced Density Matrices};
  Cioslowski,~J., Ed.; Springer US: Boston, MA, 2000; pp 33--56\relax
\mciteBstWouldAddEndPuncttrue
\mciteSetBstMidEndSepPunct{\mcitedefaultmidpunct}
{\mcitedefaultendpunct}{\mcitedefaultseppunct}\relax
\EndOfBibitem
\bibitem[Herbert and Harriman(2007)Herbert, and Harriman]{Herbert:2007p261}
Herbert,~J.~M.; Harriman,~J.~E. {Cumulants, Extensivity, and the Connected
  Formulation of the Contracted Schr{\"o}dinger Equation}. \emph{Adv. Chem.
  Phys.} \textbf{2007}, \emph{134}, 261\relax
\mciteBstWouldAddEndPuncttrue
\mciteSetBstMidEndSepPunct{\mcitedefaultmidpunct}
{\mcitedefaultendpunct}{\mcitedefaultseppunct}\relax
\EndOfBibitem
\bibitem[Kong and Valeev(2011)Kong, and Valeev]{Kong:2011p214109}
Kong,~L.; Valeev,~E.~F. {A novel interpretation of reduced density matrix and
  cumulant for electronic structure theories}. \emph{J. Chem. Phys.}
  \textbf{2011}, \emph{134}, 214109--214109--9\relax
\mciteBstWouldAddEndPuncttrue
\mciteSetBstMidEndSepPunct{\mcitedefaultmidpunct}
{\mcitedefaultendpunct}{\mcitedefaultseppunct}\relax
\EndOfBibitem
\bibitem[Hanauer and K{\"o}hn(2012)Hanauer, and K{\"o}hn]{Hanauer:2012p50}
Hanauer,~M.; K{\"o}hn,~A. {Meaning and magnitude of the reduced density matrix
  cumulants}. \emph{Chem. Phys.} \textbf{2012}, \emph{401}, 50--61\relax
\mciteBstWouldAddEndPuncttrue
\mciteSetBstMidEndSepPunct{\mcitedefaultmidpunct}
{\mcitedefaultendpunct}{\mcitedefaultseppunct}\relax
\EndOfBibitem
\bibitem[Colmenero and Valdemoro(1993)Colmenero, and
  Valdemoro]{Colmenero:1993p979}
Colmenero,~F.; Valdemoro,~C. {Approximating q-order reduced density matrices in
  terms of the lower-order ones. II. Applications}. \emph{Phys. Rev. A}
  \textbf{1993}, \emph{47}, 979--985\relax
\mciteBstWouldAddEndPuncttrue
\mciteSetBstMidEndSepPunct{\mcitedefaultmidpunct}
{\mcitedefaultendpunct}{\mcitedefaultseppunct}\relax
\EndOfBibitem
\bibitem[Nakatsuji and Yasuda(1996)Nakatsuji, and Yasuda]{Nakatsuji:1996p1039}
Nakatsuji,~H.; Yasuda,~K. {Direct Determination of the Quantum-Mechanical
  Density Matrix Using the Density Equation}. \emph{Phys. Rev. Lett.}
  \textbf{1996}, \emph{76}, 1039--1042\relax
\mciteBstWouldAddEndPuncttrue
\mciteSetBstMidEndSepPunct{\mcitedefaultmidpunct}
{\mcitedefaultendpunct}{\mcitedefaultseppunct}\relax
\EndOfBibitem
\bibitem[Nakata \latin{et~al.}(2001)Nakata, Nakatsuji, Ehara, Fukuda, Nakata,
  and Fujisawa]{Nakata:2001p8282}
Nakata,~M.; Nakatsuji,~H.; Ehara,~M.; Fukuda,~M.; Nakata,~K.; Fujisawa,~K.
  {Variational calculations of fermion second-order reduced density matrices by
  semidefinite programming algorithm}. \emph{J. Chem. Phys.} \textbf{2001},
  \emph{114}, 8282\relax
\mciteBstWouldAddEndPuncttrue
\mciteSetBstMidEndSepPunct{\mcitedefaultmidpunct}
{\mcitedefaultendpunct}{\mcitedefaultseppunct}\relax
\EndOfBibitem
\bibitem[Nakata \latin{et~al.}(2002)Nakata, Ehara, and
  Nakatsuji]{Nakata:2002p5432}
Nakata,~M.; Ehara,~M.; Nakatsuji,~H. {Density matrix variational theory:
  Application to the potential energy surfaces and strongly correlated
  systems}. \emph{J. Chem. Phys.} \textbf{2002}, \emph{116}, 5432--5439\relax
\mciteBstWouldAddEndPuncttrue
\mciteSetBstMidEndSepPunct{\mcitedefaultmidpunct}
{\mcitedefaultendpunct}{\mcitedefaultseppunct}\relax
\EndOfBibitem
\bibitem[Mazziotti(2006)]{Mazziotti:2006p143002}
Mazziotti,~D.~A. {Anti-Hermitian Contracted Schr{\"o}dinger Equation: Direct
  Determination of the Two-Electron Reduced Density Matrices of Many-Electron
  Molecules}. \emph{Phys. Rev. Lett.} \textbf{2006}, \emph{97}, 143002\relax
\mciteBstWouldAddEndPuncttrue
\mciteSetBstMidEndSepPunct{\mcitedefaultmidpunct}
{\mcitedefaultendpunct}{\mcitedefaultseppunct}\relax
\EndOfBibitem
\bibitem[Kollmar(2006)]{Kollmar:2006p084108}
Kollmar,~C. {A size extensive energy functional derived from a double
  configuration interaction approach: The role of N representability
  conditions}. \emph{J. Chem. Phys.} \textbf{2006}, \emph{125}, 084108\relax
\mciteBstWouldAddEndPuncttrue
\mciteSetBstMidEndSepPunct{\mcitedefaultmidpunct}
{\mcitedefaultendpunct}{\mcitedefaultseppunct}\relax
\EndOfBibitem
\bibitem[DePrince and Mazziotti(2007)DePrince, and
  Mazziotti]{DePrince:2007p042501}
DePrince,~A.~E.; Mazziotti,~D.~A. {Parametric approach to variational
  two-electron reduced-density-matrix theory}. \emph{Phys. Rev. A}
  \textbf{2007}, \emph{76}, 042501\relax
\mciteBstWouldAddEndPuncttrue
\mciteSetBstMidEndSepPunct{\mcitedefaultmidpunct}
{\mcitedefaultendpunct}{\mcitedefaultseppunct}\relax
\EndOfBibitem
\bibitem[DePrince(2016)]{DePrince:2016p164109}
DePrince,~A.~E. {Variational optimization of the two-electron reduced-density
  matrix under pure-state N-representability conditions}. \emph{J. Chem. Phys.}
  \textbf{2016}, \emph{145}, 164109\relax
\mciteBstWouldAddEndPuncttrue
\mciteSetBstMidEndSepPunct{\mcitedefaultmidpunct}
{\mcitedefaultendpunct}{\mcitedefaultseppunct}\relax
\EndOfBibitem
\bibitem[Mazziotti(2008)]{Mazziotti:2008p253002}
Mazziotti,~D.~A. {Parametrization of the Two-Electron Reduced Density Matrix
  for its Direct Calculation without the Many-Electron Wave Function}.
  \emph{Phys. Rev. Lett.} \textbf{2008}, \emph{101}, 253002\relax
\mciteBstWouldAddEndPuncttrue
\mciteSetBstMidEndSepPunct{\mcitedefaultmidpunct}
{\mcitedefaultendpunct}{\mcitedefaultseppunct}\relax
\EndOfBibitem
\bibitem[Mazziotti(2010)]{Mazziotti:2010p062515}
Mazziotti,~D.~A. {Parametrization of the two-electron reduced density matrix
  for its direct calculation without the many-electron wave function:
  Generalizations and applications}. \emph{Phys. Rev. A} \textbf{2010},
  \emph{81}, 062515\relax
\mciteBstWouldAddEndPuncttrue
\mciteSetBstMidEndSepPunct{\mcitedefaultmidpunct}
{\mcitedefaultendpunct}{\mcitedefaultseppunct}\relax
\EndOfBibitem
\bibitem[DePrince and Mazziotti(2012)DePrince, and
  Mazziotti]{DePrince:2012p1917}
DePrince,~A.~E.; Mazziotti,~D.~A. {Connection of an elementary class of
  parametric two-electron reduced-density-matrix methods to the coupled
  electron-pair approximations}. \emph{Mol. Phys.} \textbf{2012}, \emph{110},
  1917--1925\relax
\mciteBstWouldAddEndPuncttrue
\mciteSetBstMidEndSepPunct{\mcitedefaultmidpunct}
{\mcitedefaultendpunct}{\mcitedefaultseppunct}\relax
\EndOfBibitem
\bibitem[Kutzelnigg(1991)]{Kutzelnigg:1991p349}
Kutzelnigg,~W. {Error analysis and improvements of coupled-cluster theory}.
  \emph{Theor. Chem. Acc.} \textbf{1991}, \emph{80}, 349--386\relax
\mciteBstWouldAddEndPuncttrue
\mciteSetBstMidEndSepPunct{\mcitedefaultmidpunct}
{\mcitedefaultendpunct}{\mcitedefaultseppunct}\relax
\EndOfBibitem
\bibitem[Kutzelnigg(1998)]{Kutzelnigg:1998p65}
Kutzelnigg,~W. {Almost variational coupled cluster theory}. \emph{Mol. Phys.}
  \textbf{1998}, \emph{94}, 65--71\relax
\mciteBstWouldAddEndPuncttrue
\mciteSetBstMidEndSepPunct{\mcitedefaultmidpunct}
{\mcitedefaultendpunct}{\mcitedefaultseppunct}\relax
\EndOfBibitem
\bibitem[Van~Voorhis and Head-Gordon(2000)Van~Voorhis, and
  Head-Gordon]{VanVoorhis:2000p8873}
Van~Voorhis,~T.; Head-Gordon,~M. {Benchmark variational coupled cluster doubles
  results}. \emph{J. Chem. Phys.} \textbf{2000}, \emph{113}, 8873\relax
\mciteBstWouldAddEndPuncttrue
\mciteSetBstMidEndSepPunct{\mcitedefaultmidpunct}
{\mcitedefaultendpunct}{\mcitedefaultseppunct}\relax
\EndOfBibitem
\bibitem[Kutzelnigg(1982)]{Kutzelnigg:1982p3081}
Kutzelnigg,~W. {Quantum chemistry in Fock space. I. The universal wave and
  energy operators}. \emph{J. Chem. Phys.} \textbf{1982}, \emph{77},
  3081--3097\relax
\mciteBstWouldAddEndPuncttrue
\mciteSetBstMidEndSepPunct{\mcitedefaultmidpunct}
{\mcitedefaultendpunct}{\mcitedefaultseppunct}\relax
\EndOfBibitem
\bibitem[Bartlett \latin{et~al.}(1989)Bartlett, Kucharski, and
  Noga]{Bartlett:1989p133}
Bartlett,~R.~J.; Kucharski,~S.~A.; Noga,~J. {Alternative coupled-cluster
  ans{\"a}tze II. The unitary coupled-cluster method}. \emph{Chem. Phys. Lett.}
  \textbf{1989}, \emph{155}, 133--140\relax
\mciteBstWouldAddEndPuncttrue
\mciteSetBstMidEndSepPunct{\mcitedefaultmidpunct}
{\mcitedefaultendpunct}{\mcitedefaultseppunct}\relax
\EndOfBibitem
\bibitem[Watts \latin{et~al.}(1989)Watts, Trucks, and Bartlett]{Watts:1989p359}
Watts,~J.~D.; Trucks,~G.~W.; Bartlett,~R.~J. {The unitary coupled-cluster
  approach and molecular properties. Applications of the UCC(4) method}.
  \emph{Chem. Phys. Lett.} \textbf{1989}, \emph{157}, 359--366\relax
\mciteBstWouldAddEndPuncttrue
\mciteSetBstMidEndSepPunct{\mcitedefaultmidpunct}
{\mcitedefaultendpunct}{\mcitedefaultseppunct}\relax
\EndOfBibitem
\bibitem[Szalay \latin{et~al.}(1995)Szalay, Nooijen, and
  Bartlett]{Szalay:1995p281}
Szalay,~P.~G.; Nooijen,~M.; Bartlett,~R.~J. {Alternative ans{\"a}tze in single
  reference coupled-cluster theory. III. A critical analysis of different
  methods}. \emph{J. Chem. Phys.} \textbf{1995}, \emph{103}, 281--298\relax
\mciteBstWouldAddEndPuncttrue
\mciteSetBstMidEndSepPunct{\mcitedefaultmidpunct}
{\mcitedefaultendpunct}{\mcitedefaultseppunct}\relax
\EndOfBibitem
\bibitem[Cooper and Knowles(2010)Cooper, and Knowles]{Cooper:2010p234102}
Cooper,~B.; Knowles,~P.~J. {Benchmark studies of variational, unitary and
  extended coupled cluster methods}. \emph{J. Chem. Phys.} \textbf{2010},
  \emph{133}, 234102\relax
\mciteBstWouldAddEndPuncttrue
\mciteSetBstMidEndSepPunct{\mcitedefaultmidpunct}
{\mcitedefaultendpunct}{\mcitedefaultseppunct}\relax
\EndOfBibitem
\bibitem[Evangelista(2011)]{Evangelista:2011p224102}
Evangelista,~F.~A. {Alternative single-reference coupled cluster approaches for
  multireference problems: The simpler, the better}. \emph{J. Chem. Phys.}
  \textbf{2011}, \emph{134}, 224102--224102--13\relax
\mciteBstWouldAddEndPuncttrue
\mciteSetBstMidEndSepPunct{\mcitedefaultmidpunct}
{\mcitedefaultendpunct}{\mcitedefaultseppunct}\relax
\EndOfBibitem
\bibitem[Nakata and Yasuda(2009)Nakata, and Yasuda]{Nakata:2009p042109}
Nakata,~M.; Yasuda,~K. {Size extensivity of the variational
  reduced-density-matrix method}. \emph{Phys. Rev. A} \textbf{2009}, \emph{80},
  042109\relax
\mciteBstWouldAddEndPuncttrue
\mciteSetBstMidEndSepPunct{\mcitedefaultmidpunct}
{\mcitedefaultendpunct}{\mcitedefaultseppunct}\relax
\EndOfBibitem
\bibitem[Van~Aggelen \latin{et~al.}(2010)Van~Aggelen, Verstichel, Bultinck,
  Van~Neck, Ayers, and Cooper]{vanAggelen:2010p114112}
Van~Aggelen,~H.; Verstichel,~B.; Bultinck,~P.; Van~Neck,~D.; Ayers,~P.~W.;
  Cooper,~D.~L. {Chemical verification of variational second-order density
  matrix based potential energy surfaces for the N[sub 2] isoelectronic
  series}. \emph{J. Chem. Phys.} \textbf{2010}, \emph{132}, 114112\relax
\mciteBstWouldAddEndPuncttrue
\mciteSetBstMidEndSepPunct{\mcitedefaultmidpunct}
{\mcitedefaultendpunct}{\mcitedefaultseppunct}\relax
\EndOfBibitem
\bibitem[Verstichel \latin{et~al.}(2010)Verstichel, Van~Aggelen, Van~Neck,
  Ayers, and Bultinck]{Verstichel:2010p114113}
Verstichel,~B.; Van~Aggelen,~H.; Van~Neck,~D.; Ayers,~P.~W.; Bultinck,~P.
  {Subsystem constraints in variational second order density matrix
  optimization: Curing the dissociative behavior}. \emph{J. Chem. Phys.}
  \textbf{2010}, \emph{132}, 114113\relax
\mciteBstWouldAddEndPuncttrue
\mciteSetBstMidEndSepPunct{\mcitedefaultmidpunct}
{\mcitedefaultendpunct}{\mcitedefaultseppunct}\relax
\EndOfBibitem
\bibitem[Scheiner \latin{et~al.}(1987)Scheiner, Scuseria, Rice, Lee, and
  Schaefer]{Scheiner:1987p5361}
Scheiner,~A.~C.; Scuseria,~G.~E.; Rice,~J.~E.; Lee,~T.~J.; Schaefer,~H.~F.
  {Analytic evaluation of energy gradients for the single and double excitation
  coupled cluster (CCSD) wave function: Theory and application}. \emph{J. Chem.
  Phys.} \textbf{1987}, \emph{87}, 5361--5373\relax
\mciteBstWouldAddEndPuncttrue
\mciteSetBstMidEndSepPunct{\mcitedefaultmidpunct}
{\mcitedefaultendpunct}{\mcitedefaultseppunct}\relax
\EndOfBibitem
\bibitem[Salter \latin{et~al.}(1989)Salter, Trucks, and
  Bartlett]{Salter:1989p1752}
Salter,~E.~A.; Trucks,~G.~W.; Bartlett,~R.~J. {Analytic energy derivatives in
  many-body methods. I. First derivatives}. \emph{J. Chem. Phys.}
  \textbf{1989}, \emph{90}, 1752--1766\relax
\mciteBstWouldAddEndPuncttrue
\mciteSetBstMidEndSepPunct{\mcitedefaultmidpunct}
{\mcitedefaultendpunct}{\mcitedefaultseppunct}\relax
\EndOfBibitem
\bibitem[Gauss \latin{et~al.}(1991)Gauss, Stanton, and
  Bartlett]{Gauss:1991p2623}
Gauss,~J.; Stanton,~J.~F.; Bartlett,~R.~J. {Coupled-cluster open-shell analytic
  gradients: Implementation of the direct product decomposition approach in
  energy gradient calculations}. \emph{J. Chem. Phys.} \textbf{1991},
  \emph{95}, 2623\relax
\mciteBstWouldAddEndPuncttrue
\mciteSetBstMidEndSepPunct{\mcitedefaultmidpunct}
{\mcitedefaultendpunct}{\mcitedefaultseppunct}\relax
\EndOfBibitem
\bibitem[Gauss \latin{et~al.}(1991)Gauss, Lauderdale, Stanton, Watts, and
  Bartlett]{Gauss:1991p207}
Gauss,~J.; Lauderdale,~W.~J.; Stanton,~J.~F.; Watts,~J.~D.; Bartlett,~R.~J.
  {Analytic energy gradients for open-shell coupled-cluster singles and doubles
  (CCSD) calculations using restricted open-shell Hartree{\textemdash}Fock
  (ROHF) reference functions}. \emph{Chem. Phys. Lett.} \textbf{1991},
  \emph{182}, 207--215\relax
\mciteBstWouldAddEndPuncttrue
\mciteSetBstMidEndSepPunct{\mcitedefaultmidpunct}
{\mcitedefaultendpunct}{\mcitedefaultseppunct}\relax
\EndOfBibitem
\bibitem[Copan \latin{et~al.}(2014)Copan, Sokolov, and
  Schaefer]{Copan:2014p2389}
Copan,~A.~V.; Sokolov,~A.~Y.; Schaefer,~H.~F. {Benchmark Study of Density
  Cumulant Functional Theory: Thermochemistry and Kinetics}. \emph{J. Chem.
  Theory Comput.} \textbf{2014}, \emph{10}, 2389--2398\relax
\mciteBstWouldAddEndPuncttrue
\mciteSetBstMidEndSepPunct{\mcitedefaultmidpunct}
{\mcitedefaultendpunct}{\mcitedefaultseppunct}\relax
\EndOfBibitem
\bibitem[Mullinax \latin{et~al.}(2015)Mullinax, Sokolov, and
  Schaefer]{Mullinax:2015p2487}
Mullinax,~J.~W.; Sokolov,~A.~Y.; Schaefer,~H.~F. {Can Density Cumulant
  Functional Theory Describe Static Correlation Effects?} \emph{J. Chem. Theory
  Comput.} \textbf{2015}, \emph{11}, 2487--2495\relax
\mciteBstWouldAddEndPuncttrue
\mciteSetBstMidEndSepPunct{\mcitedefaultmidpunct}
{\mcitedefaultendpunct}{\mcitedefaultseppunct}\relax
\EndOfBibitem
\bibitem[Bozkaya and Sherrill(2013)Bozkaya, and Sherrill]{Bozkaya:2013p054104}
Bozkaya,~U.; Sherrill,~C.~D. {Orbital-optimized coupled-electron pair theory
  and its analytic gradients: Accurate equilibrium geometries, harmonic
  vibrational frequencies, and hydrogen transfer reactions}. \emph{J. Chem.
  Phys.} \textbf{2013}, \emph{139}, 054104--054104--12\relax
\mciteBstWouldAddEndPuncttrue
\mciteSetBstMidEndSepPunct{\mcitedefaultmidpunct}
{\mcitedefaultendpunct}{\mcitedefaultseppunct}\relax
\EndOfBibitem
\bibitem[Norman(2011)]{Norman:2011p20519}
Norman,~P. {A Perspective on Nonresonant and Resonant Electronic Response
  Theory for Time-Dependent Molecular Properties}. \emph{Phys. Chem. Chem.
  Phys.} \textbf{2011}, \emph{13}, 20519--20535\relax
\mciteBstWouldAddEndPuncttrue
\mciteSetBstMidEndSepPunct{\mcitedefaultmidpunct}
{\mcitedefaultendpunct}{\mcitedefaultseppunct}\relax
\EndOfBibitem
\bibitem[Helgaker \latin{et~al.}(2012)Helgaker, Coriani, J{\o}rgensen,
  Kristensen, Olsen, and Ruud]{Helgaker:2012p543}
Helgaker,~T.; Coriani,~S.; J{\o}rgensen,~P.; Kristensen,~K.; Olsen,~J.;
  Ruud,~K. {Recent Advances in Wave Function-Based Methods of
  Molecular-Property Calculations}. \emph{Chem. Rev.} \textbf{2012},
  \emph{112}, 543--631\relax
\mciteBstWouldAddEndPuncttrue
\mciteSetBstMidEndSepPunct{\mcitedefaultmidpunct}
{\mcitedefaultendpunct}{\mcitedefaultseppunct}\relax
\EndOfBibitem
\bibitem[Kristensen \latin{et~al.}(2009)Kristensen, Kauczor, Kj{\ae}rgaard, and
  J{\o}rgensen]{Kristensen:2009p044112}
Kristensen,~K.; Kauczor,~J.; Kj{\ae}rgaard,~T.; J{\o}rgensen,~P. {Quasienergy
  Formulation of Damped Response Theory}. \emph{J. Chem. Phys.} \textbf{2009},
  \emph{131}, 044112\relax
\mciteBstWouldAddEndPuncttrue
\mciteSetBstMidEndSepPunct{\mcitedefaultmidpunct}
{\mcitedefaultendpunct}{\mcitedefaultseppunct}\relax
\EndOfBibitem
\bibitem[Olsen and J{\o}rgensen(1985)Olsen, and J{\o}rgensen]{Olsen:1985p3235}
Olsen,~J.; J{\o}rgensen,~P. {Linear and Nonlinear Response Functions for an
  Exact State and for an MCSCF State}. \emph{J. Chem. Phys.} \textbf{1985},
  \emph{82}, 3235--3264\relax
\mciteBstWouldAddEndPuncttrue
\mciteSetBstMidEndSepPunct{\mcitedefaultmidpunct}
{\mcitedefaultendpunct}{\mcitedefaultseppunct}\relax
\EndOfBibitem
\bibitem[Sauer(2011)]{Sauer:2011}
Sauer,~S. P.~A. \emph{{Molecular Electromagnetism: A Computational Chemistry
  Approach}}; Oxford University Press: Oxford, 2011\relax
\mciteBstWouldAddEndPuncttrue
\mciteSetBstMidEndSepPunct{\mcitedefaultmidpunct}
{\mcitedefaultendpunct}{\mcitedefaultseppunct}\relax
\EndOfBibitem
\bibitem[Parrish \latin{et~al.}(2017)Parrish, Burns, Smith, Simmonett,
  DePrince, Hohenstein, Bozkaya, Sokolov, Di~Remigio, Richard, Gonthier, James,
  McAlexander, Kumar, Saitow, Wang, Pritchard, Verma, Schaefer, Patkowski,
  King, Valeev, Evangelista, Turney, Crawford, and Sherrill]{Parrish:2017p3185}
Parrish,~R.~M.; Burns,~L.~A.; Smith,~D. G.~A.; Simmonett,~A.~C.;
  DePrince,~A.~E.; Hohenstein,~E.~G.; Bozkaya,~U.; Sokolov,~A.~Y.;
  Di~Remigio,~R.; Richard,~R.~M.; Gonthier,~J.~F.; James,~A.~M.;
  McAlexander,~H.~R.; Kumar,~A.; Saitow,~M.; Wang,~X.; Pritchard,~B.~P.;
  Verma,~P.; Schaefer,~H.~F.; Patkowski,~K.; King,~R.~A.; Valeev,~E.~F.;
  Evangelista,~F.~A.; Turney,~J.~M.; Crawford,~T.~D.; Sherrill,~C.~D. {Psi41.1:
  An Open-Source Electronic Structure Program Emphasizing Automation, Advanced
  Libraries, and Interoperability}. \emph{J. Chem. Theory Comput.}
  \textbf{2017}, \emph{13}, 3185--3197\relax
\mciteBstWouldAddEndPuncttrue
\mciteSetBstMidEndSepPunct{\mcitedefaultmidpunct}
{\mcitedefaultendpunct}{\mcitedefaultseppunct}\relax
\EndOfBibitem
\bibitem[Sun \latin{et~al.}(2018)Sun, Berkelbach, Blunt, Booth, Guo, Li, Liu,
  McClain, Sayfutyarova, Sharma, Wouters, and Chan]{Sun:2018pe1340}
Sun,~Q.; Berkelbach,~T.~C.; Blunt,~N.~S.; Booth,~G.~H.; Guo,~S.; Li,~Z.;
  Liu,~J.; McClain,~J.~D.; Sayfutyarova,~E.~R.; Sharma,~S.; Wouters,~S.;
  Chan,~G. K.-L. {PySCF: the Python-based simulations of chemistry framework}.
  \emph{WIREs Comput. Mol. Sci.} \textbf{2018}, \emph{8}, e1340\relax
\mciteBstWouldAddEndPuncttrue
\mciteSetBstMidEndSepPunct{\mcitedefaultmidpunct}
{\mcitedefaultendpunct}{\mcitedefaultseppunct}\relax
\EndOfBibitem
\bibitem[Davidson(1975)]{Davidson:1975p87}
Davidson,~E.~R. {The Iterative Calculation of a Few of the Lowest Eigenvalues
  and Corresponding Eigenvectors of Large Real-Symmetric Matrices}. \emph{J.
  Comput. Phys.} \textbf{1975}, \emph{17}, 87--94\relax
\mciteBstWouldAddEndPuncttrue
\mciteSetBstMidEndSepPunct{\mcitedefaultmidpunct}
{\mcitedefaultendpunct}{\mcitedefaultseppunct}\relax
\EndOfBibitem
\bibitem[Liu(1978)]{Liu:1978p49}
Liu,~B. \emph{{The Simultaneous Expansion Method for the Iterative Solution of
  Several of the Lowest-Lying Eigenvalues and Corresponding Eigenvectors of
  Large Real-Symmetric Matrices}}; 1978; pp 49--53\relax
\mciteBstWouldAddEndPuncttrue
\mciteSetBstMidEndSepPunct{\mcitedefaultmidpunct}
{\mcitedefaultendpunct}{\mcitedefaultseppunct}\relax
\EndOfBibitem
\bibitem[Shao \latin{et~al.}(2015)Shao, Gan, Epifanovsky, Gilbert, Wormit,
  Kussmann, Lange, Behn, Deng, Feng, Ghosh, Goldey, Horn, Jacobson, Kaliman,
  Khaliullin, Kuś, Landau, Liu, Proynov, Rhee, Richard, Rohrdanz, Steele,
  Sundstrom, III, Zimmerman, Zuev, Albrecht, Alguire, Austin, Beran, Bernard,
  Berquist, Brandhorst, Bravaya, Brown, Casanova, Chang, Chen, Chien, Closser,
  Crittenden, Diedenhofen, Jr., Do, Dutoi, Edgar, Fatehi, Fusti-Molnar,
  Ghysels, Golubeva-Zadorozhnaya, Gomes, Hanson-Heine, Harbach, Hauser,
  Hohenstein, Holden, Jagau, Ji, Kaduk, Khistyaev, Kim, Kim, King, Klunzinger,
  Kosenkov, Kowalczyk, Krauter, Lao, Laurent, Lawler, Levchenko, Lin, Liu,
  Livshits, Lochan, Luenser, Manohar, Manzer, Mao, Mardirossian, Marenich,
  Maurer, Mayhall, Neuscamman, Oana, Olivares-Amaya, O’Neill, Parkhill,
  Perrine, Peverati, Prociuk, Rehn, Rosta, Russ, Sharada, Sharma, Small, Sodt,
  Stein, Stück, Su, Thom, Tsuchimochi, Vanovschi, Vogt, Vydrov, Wang, Watson,
  Wenzel, White, Williams, Yang, Yeganeh, Yost, You, Zhang, Zhang, Zhao,
  Brooks, Chan, Chipman, Cramer, III, Gordon, Hehre, Klamt, III, Schmidt,
  Sherrill, Truhlar, Warshel, Xu, Aspuru-Guzik, Baer, Bell, Besley, Chai,
  Dreuw, Dunietz, Furlani, Gwaltney, Hsu, Jung, Kong, Lambrecht, Liang,
  Ochsenfeld, Rassolov, Slipchenko, Subotnik, Voorhis, Herbert, Krylov, Gill,
  and Head-Gordon]{qchem:44}
Shao,~Y.; Gan,~Z.; Epifanovsky,~E.; Gilbert,~A.~T.; Wormit,~M.; Kussmann,~J.;
  Lange,~A.~W.; Behn,~A.; Deng,~J.; Feng,~X.; Ghosh,~D.; Goldey,~M.;
  Horn,~P.~R.; Jacobson,~L.~D.; Kaliman,~I.; Khaliullin,~R.~Z.; Kuś,~T.;
  Landau,~A.; Liu,~J.; Proynov,~E.~I.; Rhee,~Y.~M.; Richard,~R.~M.;
  Rohrdanz,~M.~A.; Steele,~R.~P.; Sundstrom,~E.~J.; III,~H. L.~W.;
  Zimmerman,~P.~M.; Zuev,~D.; Albrecht,~B.; Alguire,~E.; Austin,~B.; Beran,~G.
  J.~O.; Bernard,~Y.~A.; Berquist,~E.; Brandhorst,~K.; Bravaya,~K.~B.;
  Brown,~S.~T.; Casanova,~D.; Chang,~C.-M.; Chen,~Y.; Chien,~S.~H.;
  Closser,~K.~D.; Crittenden,~D.~L.; Diedenhofen,~M.; Jr.,~R. A.~D.; Do,~H.;
  Dutoi,~A.~D.; Edgar,~R.~G.; Fatehi,~S.; Fusti-Molnar,~L.; Ghysels,~A.;
  Golubeva-Zadorozhnaya,~A.; Gomes,~J.; Hanson-Heine,~M.~W.; Harbach,~P.~H.;
  Hauser,~A.~W.; Hohenstein,~E.~G.; Holden,~Z.~C.; Jagau,~T.-C.; Ji,~H.;
  Kaduk,~B.; Khistyaev,~K.; Kim,~J.; Kim,~J.; King,~R.~A.; Klunzinger,~P.;
  Kosenkov,~D.; Kowalczyk,~T.; Krauter,~C.~M.; Lao,~K.~U.; Laurent,~A.~D.;
  Lawler,~K.~V.; Levchenko,~S.~V.; Lin,~C.~Y.; Liu,~F.; Livshits,~E.;
  Lochan,~R.~C.; Luenser,~A.; Manohar,~P.; Manzer,~S.~F.; Mao,~S.-P.;
  Mardirossian,~N.; Marenich,~A.~V.; Maurer,~S.~A.; Mayhall,~N.~J.;
  Neuscamman,~E.; Oana,~C.~M.; Olivares-Amaya,~R.; O’Neill,~D.~P.;
  Parkhill,~J.~A.; Perrine,~T.~M.; Peverati,~R.; Prociuk,~A.; Rehn,~D.~R.;
  Rosta,~E.; Russ,~N.~J.; Sharada,~S.~M.; Sharma,~S.; Small,~D.~W.; Sodt,~A.;
  Stein,~T.; Stück,~D.; Su,~Y.-C.; Thom,~A.~J.; Tsuchimochi,~T.;
  Vanovschi,~V.; Vogt,~L.; Vydrov,~O.; Wang,~T.; Watson,~M.~A.; Wenzel,~J.;
  White,~A.; Williams,~C.~F.; Yang,~J.; Yeganeh,~S.; Yost,~S.~R.; You,~Z.-Q.;
  Zhang,~I.~Y.; Zhang,~X.; Zhao,~Y.; Brooks,~B.~R.; Chan,~G.~K.;
  Chipman,~D.~M.; Cramer,~C.~J.; III,~W. A.~G.; Gordon,~M.~S.; Hehre,~W.~J.;
  Klamt,~A.; III,~H. F.~S.; Schmidt,~M.~W.; Sherrill,~C.~D.; Truhlar,~D.~G.;
  Warshel,~A.; Xu,~X.; Aspuru-Guzik,~A.; Baer,~R.; Bell,~A.~T.; Besley,~N.~A.;
  Chai,~J.-D.; Dreuw,~A.; Dunietz,~B.~D.; Furlani,~T.~R.; Gwaltney,~S.~R.;
  Hsu,~C.-P.; Jung,~Y.; Kong,~J.; Lambrecht,~D.~S.; Liang,~W.; Ochsenfeld,~C.;
  Rassolov,~V.~A.; Slipchenko,~L.~V.; Subotnik,~J.~E.; Voorhis,~T.~V.;
  Herbert,~J.~M.; Krylov,~A.~I.; Gill,~P.~M.; Head-Gordon,~M. Advances in
  molecular quantum chemistry contained in the Q-Chem 4 program package.
  \emph{Mol. Phys.} \textbf{2015}, \emph{113}, 184--215\relax
\mciteBstWouldAddEndPuncttrue
\mciteSetBstMidEndSepPunct{\mcitedefaultmidpunct}
{\mcitedefaultendpunct}{\mcitedefaultseppunct}\relax
\EndOfBibitem
\bibitem[K\'allay \latin{et~al.}()K\'allay, Rolik, Csontos, Nagy, Samu, Mester,
  Cs\'oka, Szab\'o, Ladjánszki, Szegedy, Lad\'oczki, Petrov, Farkas, Mezei,
  and Hégely.]{MRCC}
K\'allay,~M.; Rolik,~Z.; Csontos,~J.; Nagy,~P.; Samu,~G.; Mester,~D.;
  Cs\'oka,~J.; Szab\'o,~B.; Ladjánszki,~I.; Szegedy,~L.; Lad\'oczki,~B.;
  Petrov,~K.; Farkas,~M.; Mezei,~P.~D.; Hégely.,~B. MRCC, a quantum chemical
  program suite. See also: Z. Rolik, L. Szegedy, I. Ladj\'anszki, B.
  Lad\'oczki, and M. K\'allay, J. Chem. Phys. 139, 094105 (2013), as well as:
  www.mrcc.hu.\relax
\mciteBstWouldAddEndPunctfalse
\mciteSetBstMidEndSepPunct{\mcitedefaultmidpunct}
{}{\mcitedefaultseppunct}\relax
\EndOfBibitem
\bibitem[Kendall \latin{et~al.}(1992)Kendall, Dunning~Jr, and
  Harrison]{Kendall:1992p6796}
Kendall,~R.~A.; Dunning~Jr,~T.~H.; Harrison,~R.~J. {Electron affinities of the
  first-row atoms revisited. Systematic basis sets and wave functions}.
  \emph{J. Chem. Phys.} \textbf{1992}, \emph{96}, 6796--6806\relax
\mciteBstWouldAddEndPuncttrue
\mciteSetBstMidEndSepPunct{\mcitedefaultmidpunct}
{\mcitedefaultendpunct}{\mcitedefaultseppunct}\relax
\EndOfBibitem
\bibitem[Widmark \latin{et~al.}(1990)Widmark, Malmqvist, and
  Roos]{Widmark:1990p291}
Widmark,~P.-O.; Malmqvist,~P.~{\AA}.; Roos,~B.~O. {Density matrix averaged
  atomic natural orbital (ANO) basis sets for correlated molecular wave
  functions}. \emph{Theoret. Chim. Acta} \textbf{1990}, \emph{77},
  291--306\relax
\mciteBstWouldAddEndPuncttrue
\mciteSetBstMidEndSepPunct{\mcitedefaultmidpunct}
{\mcitedefaultendpunct}{\mcitedefaultseppunct}\relax
\EndOfBibitem
\bibitem[Daday \latin{et~al.}(2012)Daday, Smart, Booth, Alavi, and
  Filippi]{Daday:2012p4441}
Daday,~C.; Smart,~S.; Booth,~G.~H.; Alavi,~A.; Filippi,~C. {Full Configuration
  Interaction Excitations of Ethene and Butadiene: Resolution of an Ancient
  Question}. \emph{J. Chem. Theory Comput.} \textbf{2012}, \emph{8},
  4441--4451\relax
\mciteBstWouldAddEndPuncttrue
\mciteSetBstMidEndSepPunct{\mcitedefaultmidpunct}
{\mcitedefaultendpunct}{\mcitedefaultseppunct}\relax
\EndOfBibitem
\bibitem[Zimmerman(2017)]{Zimmerman:2017p4712}
Zimmerman,~P.~M. {Singlet{\textendash}Triplet Gaps through Incremental Full
  Configuration Interaction}. \emph{J. Phys. Chem. A} \textbf{2017},
  \emph{121}, 4712--4720\relax
\mciteBstWouldAddEndPuncttrue
\mciteSetBstMidEndSepPunct{\mcitedefaultmidpunct}
{\mcitedefaultendpunct}{\mcitedefaultseppunct}\relax
\EndOfBibitem
\bibitem[Chien \latin{et~al.}(2018)Chien, Holmes, Otten, Umrigar, Sharma, and
  Zimmerman]{Chien:2018p2714}
Chien,~A.~D.; Holmes,~A.~A.; Otten,~M.; Umrigar,~C.~J.; Sharma,~S.;
  Zimmerman,~P.~M. {Excited States of Methylene, Polyenes, and Ozone from
  Heat-Bath Configuration Interaction}. \emph{J. Phys. Chem. A} \textbf{2018},
  \emph{122}, 2714--2722\relax
\mciteBstWouldAddEndPuncttrue
\mciteSetBstMidEndSepPunct{\mcitedefaultmidpunct}
{\mcitedefaultendpunct}{\mcitedefaultseppunct}\relax
\EndOfBibitem
\bibitem[Tavan and Schulten(1986)Tavan, and Schulten]{Tavan:1986p6602}
Tavan,~P.; Schulten,~K. {The low-lying electronic excitations in long polyenes:
  A PPP-MRD-CI study}. \emph{J. Chem. Phys.} \textbf{1986}, \emph{85},
  6602--6609\relax
\mciteBstWouldAddEndPuncttrue
\mciteSetBstMidEndSepPunct{\mcitedefaultmidpunct}
{\mcitedefaultendpunct}{\mcitedefaultseppunct}\relax
\EndOfBibitem
\bibitem[Tavan and Schulten(1987)Tavan, and Schulten]{Tavan:1987p4337}
Tavan,~P.; Schulten,~K. {Electronic excitations in finite and infinite
  polyenes}. \emph{Phys. Rev. B} \textbf{1987}, \emph{36}, 4337--4358\relax
\mciteBstWouldAddEndPuncttrue
\mciteSetBstMidEndSepPunct{\mcitedefaultmidpunct}
{\mcitedefaultendpunct}{\mcitedefaultseppunct}\relax
\EndOfBibitem
\bibitem[Nakayama \latin{et~al.}(1998)Nakayama, Nakano, and
  Hirao]{Nakayama:1998p157}
Nakayama,~K.; Nakano,~H.; Hirao,~K. {Theoretical study of the
  $\pi${\textrightarrow}$\pi$* excited states of linear polyenes: The energy
  gap between 11Bu+ and 21Ag{\textminus} states and their character}.
  \emph{Int. J. Quantum Chem.} \textbf{1998}, \emph{66}, 157--175\relax
\mciteBstWouldAddEndPuncttrue
\mciteSetBstMidEndSepPunct{\mcitedefaultmidpunct}
{\mcitedefaultendpunct}{\mcitedefaultseppunct}\relax
\EndOfBibitem
\bibitem[Davidson(1996)]{Davidson:1996p6161}
Davidson,~E.~R. {The Spatial Extent of the V State of Ethylene and Its Relation
  to Dynamic Correlation in the Cope Rearrangement}. \emph{J. Phys. Chem.}
  \textbf{1996}, \emph{100}, 6161--6166\relax
\mciteBstWouldAddEndPuncttrue
\mciteSetBstMidEndSepPunct{\mcitedefaultmidpunct}
{\mcitedefaultendpunct}{\mcitedefaultseppunct}\relax
\EndOfBibitem
\bibitem[Watts \latin{et~al.}(1998)Watts, Gwaltney, and
  Bartlett]{Watts:1998p6979}
Watts,~J.~D.; Gwaltney,~S.~R.; Bartlett,~R.~J. {Coupled-cluster calculations of
  the excitation energies of ethylene, butadiene, and cyclopentadiene}.
  \emph{J. Chem. Phys.} \textbf{1998}, \emph{105}, 6979--6988\relax
\mciteBstWouldAddEndPuncttrue
\mciteSetBstMidEndSepPunct{\mcitedefaultmidpunct}
{\mcitedefaultendpunct}{\mcitedefaultseppunct}\relax
\EndOfBibitem
\bibitem[M{\"u}ller \latin{et~al.}(1999)M{\"u}ller, Dallos, and
  Lischka]{Muller:1999p7176}
M{\"u}ller,~T.; Dallos,~M.; Lischka,~H. {The ethylene 1 1B1u V state
  revisited}. \emph{J. Chem. Phys.} \textbf{1999}, \emph{110}, 7176--7184\relax
\mciteBstWouldAddEndPuncttrue
\mciteSetBstMidEndSepPunct{\mcitedefaultmidpunct}
{\mcitedefaultendpunct}{\mcitedefaultseppunct}\relax
\EndOfBibitem
\bibitem[Li and Paldus(1999)Li, and Paldus]{Li:1999p177}
Li,~X.; Paldus,~J. {Size dependence of the X1Ag{\textrightarrow}11Bu excitation
  energy in linear polyenes}. \emph{Int. J. Quantum Chem.} \textbf{1999},
  \emph{74}, 177--192\relax
\mciteBstWouldAddEndPuncttrue
\mciteSetBstMidEndSepPunct{\mcitedefaultmidpunct}
{\mcitedefaultendpunct}{\mcitedefaultseppunct}\relax
\EndOfBibitem
\bibitem[Starcke \latin{et~al.}(2006)Starcke, Wormit, Schirmer, and
  Dreuw]{Starcke:2006p39}
Starcke,~J.~H.; Wormit,~M.; Schirmer,~J.; Dreuw,~A. {How much double excitation
  character do the lowest excited states of linear polyenes have?} \emph{Chem.
  Phys.} \textbf{2006}, \emph{329}, 39--49\relax
\mciteBstWouldAddEndPuncttrue
\mciteSetBstMidEndSepPunct{\mcitedefaultmidpunct}
{\mcitedefaultendpunct}{\mcitedefaultseppunct}\relax
\EndOfBibitem
\bibitem[Kurashige \latin{et~al.}(2004)Kurashige, Nakano, Nakao, and
  Hirao]{Kurashige:2004p425}
Kurashige,~Y.; Nakano,~H.; Nakao,~Y.; Hirao,~K. {The
  $\pi${\textrightarrow}$\pi$* excited states of long linear polyenes studied
  by the CASCI-MRMP method}. \emph{Chem. Phys. Lett.} \textbf{2004},
  \emph{400}, 425--429\relax
\mciteBstWouldAddEndPuncttrue
\mciteSetBstMidEndSepPunct{\mcitedefaultmidpunct}
{\mcitedefaultendpunct}{\mcitedefaultseppunct}\relax
\EndOfBibitem
\bibitem[Ghosh \latin{et~al.}(2008)Ghosh, Hachmann, Yanai, and
  Chan]{Ghosh:2008p144117}
Ghosh,~D.; Hachmann,~J.; Yanai,~T.; Chan,~G. K.-L. {Orbital optimization in the
  density matrix renormalization group, with applications to polyenes and
  $\beta$-carotene}. \emph{J. Chem. Phys.} \textbf{2008}, \emph{128},
  144117\relax
\mciteBstWouldAddEndPuncttrue
\mciteSetBstMidEndSepPunct{\mcitedefaultmidpunct}
{\mcitedefaultendpunct}{\mcitedefaultseppunct}\relax
\EndOfBibitem
\bibitem[Sokolov \latin{et~al.}(2017)Sokolov, Guo, Ronca, and
  Chan]{Sokolov:2017p244102}
Sokolov,~A.~Y.; Guo,~S.; Ronca,~E.; Chan,~G. K.-L. {Time-dependent N-electron
  valence perturbation theory with matrix product state reference wavefunctions
  for large active spaces and basis sets: Applications to the chromium dimer
  and all-trans polyenes}. \emph{J. Chem. Phys.} \textbf{2017}, \emph{146},
  244102\relax
\mciteBstWouldAddEndPuncttrue
\mciteSetBstMidEndSepPunct{\mcitedefaultmidpunct}
{\mcitedefaultendpunct}{\mcitedefaultseppunct}\relax
\EndOfBibitem
\bibitem[Schreiber \latin{et~al.}(2008)Schreiber, Silva-Junior, Sauer, and
  Thiel]{Schreiber:2008p134110}
Schreiber,~M.; Silva-Junior,~M.~R.; Sauer,~S. P.~A.; Thiel,~W. {Benchmarks for
  electronically excited states: CASPT2, CC2, CCSD, and CC3}. \emph{J. Chem.
  Phys.} \textbf{2008}, \emph{128}, 134110\relax
\mciteBstWouldAddEndPuncttrue
\mciteSetBstMidEndSepPunct{\mcitedefaultmidpunct}
{\mcitedefaultendpunct}{\mcitedefaultseppunct}\relax
\EndOfBibitem
\bibitem[Zgid \latin{et~al.}(2009)Zgid, Ghosh, Neuscamman, and
  Chan]{Zgid:2009p194107}
Zgid,~D.; Ghosh,~D.; Neuscamman,~E.; Chan,~G. K.-L. {A study of cumulant
  approximations to n-electron valence multireference perturbation theory}.
  \emph{J. Chem. Phys.} \textbf{2009}, \emph{130}, 194107\relax
\mciteBstWouldAddEndPuncttrue
\mciteSetBstMidEndSepPunct{\mcitedefaultmidpunct}
{\mcitedefaultendpunct}{\mcitedefaultseppunct}\relax
\EndOfBibitem
\bibitem[Angeli(2010)]{Angeli:2010p2436}
Angeli,~C. {An analysis of the dynamic $\sigma$ polarization in the V state of
  ethene}. \emph{Int. J. Quantum Chem.} \textbf{2010}, \emph{110},
  2436--2447\relax
\mciteBstWouldAddEndPuncttrue
\mciteSetBstMidEndSepPunct{\mcitedefaultmidpunct}
{\mcitedefaultendpunct}{\mcitedefaultseppunct}\relax
\EndOfBibitem
\bibitem[Watson and Chan(2012)Watson, and Chan]{Watson:2012p4013}
Watson,~M.~A.; Chan,~G. K.-L. {Excited States of Butadiene to Chemical
  Accuracy: Reconciling Theory and Experiment}. \emph{J. Chem. Theory Comput.}
  \textbf{2012}, \emph{8}, 4013--4018\relax
\mciteBstWouldAddEndPuncttrue
\mciteSetBstMidEndSepPunct{\mcitedefaultmidpunct}
{\mcitedefaultendpunct}{\mcitedefaultseppunct}\relax
\EndOfBibitem
\bibitem[Lindh \latin{et~al.}(2012)Lindh, Mach, and Crawford]{Lindh:2012p125}
Lindh,~G.~D.; Mach,~T.~J.; Crawford,~T.~D. {The optimized orbital coupled
  cluster doubles method and optical rotation}. \emph{Chem. Phys.}
  \textbf{2012}, \emph{401}, 125--129\relax
\mciteBstWouldAddEndPuncttrue
\mciteSetBstMidEndSepPunct{\mcitedefaultmidpunct}
{\mcitedefaultendpunct}{\mcitedefaultseppunct}\relax
\EndOfBibitem
\end{mcitethebibliography}
\end{document}